\documentclass[11pt,english,aps,nofootinbib,eqsecnum,superscriptaddress]{revtex4-2}
\usepackage[T1]{fontenc}
\usepackage[utf8]{luainputenc}
\usepackage{amsmath}
\usepackage{amssymb}
\usepackage{graphicx}

\makeatletter
\usepackage{graphicx}

\usepackage{amsmath}

\usepackage[T1]{fontenc}              
\usepackage[utf8]{luainputenc}
\usepackage{fancyhdr}
\usepackage{array}
\usepackage{appendix}
\usepackage{float}
\usepackage{wasysym}
\usepackage[marginal]{footmisc}
\usepackage{braket}
\usepackage{numprint}
\usepackage{enumitem}
\usepackage{verbatim}
\usepackage{lipsum}
\usepackage{amsfonts,amssymb,slashed}
\usepackage{scrextend}
\usepackage{setspace}
\usepackage{tikz}
\usepackage[colorlinks,linkcolor=blue,anchorcolor=blue,citecolor=blue,urlcolor=blue]{hyperref}
\usetikzlibrary{chains,patterns,patterns.meta,scopes,positioning,backgrounds,shapes,fit,shadows,calc,arrows.meta,decorations.markings,decorations.pathreplacing,decorations.pathmorphing,calligraphy}
\usepackage{tkz-euclide}
\allowdisplaybreaks[4]
\makeatother

\usepackage{babel}
\begin{document}
\title{Cutting rules for non-relativistic dark matter in solids\\ based
on Kohn--Sham orbitals }
\author{Zheng-Liang Liang}
\email{liangzl@mail.buct.edu.cn}

\affiliation{College of Mathematics and Physics, Beijing University of Chemical
Technology~\\
Beijing 100029, China}
\author{Fawei Zheng}
\email{fwzheng@bit.edu.cn}

\affiliation{Centre for Quantum Physics, Key Laboratory of Advanced Optoelectronic
Quantum Architecture and Measurement(MOE),~\\
School of Physics, Beijing Institute of Technology ~\\
Beijing, 100081, China}
\begin{abstract}
The Cutkosky cutting rules establish a direct connection between the
imaginary parts of loop amplitudes and physical observables such as
decay rates and cross sections, providing heuristic insights into
the underlying processes. This work lays a robust theoretical foundation
for the application of cutting rules in solid-state systems involving
instantaneous dark matter (DM)--electron Yukawa interaction as well
as the Coulomb potential. The cutting rules are formulated using the
single-electron wavefunctions and corresponding energy eigenvalues
obtained from the Kohn--Sham equations within density functional
theory (DFT). This framework is not only of considerable theoretical
interest but also holds significant practical relevance for studying
DM phenomenology in condensed matter systems.
\end{abstract}
\maketitle

\section{Introduction}

As direct detection techniques for \emph{dark matter} (DM) continue
to advance, theoretical studies of DM interactions with detector materials
have also become increasingly sophisticated. Particularly, the introduction
of many-body physics methods including the first-principles \textit{density
functional theory}~(DFT) method~\citep{Essig:2015cda} into the
interpretation of DM signals in semiconductor detectors (\textit{e.g}.,
SENSEI~\citep{Barak:2020fql}, DAMIC~\citep{Castello-Mor:2020jhd},
SuperCDMS~\citep{Amaral:2020ryn}, CDEX~\citep{CDEX:2022kcd} and
EDELWEISS~\citep{Arnaud:2020svb}) have spurred further discussions
on the methodology~\citep{Liang:2018bdb,Griffin:2021znd,Kahn:2021ttr,Trickle:2022fwt,Dreyer:2023ovn}
and interpretations of the DM-electron interactions~\citep{Emken:2019tni,Essig:2019xkx,Trickle:2020oki,Andersson:2020uwc,Su:2020zny,Mitridate:2021ctr,Vahsen:2021gnb,Catena:2021qsr,Chen:2022pyd,Catena:2022fnk,Li:2022acp,Wang:2023xgm}.

Due to DM-electron inelastic scattering, DM particles can excite electrons
from the valence band to the conduction band in semiconductor detectors,
thereby triggering detectable electronic signals. This process can
be equivalently interpreted as the decay of an energetic DM particle
in solids into a lower-energy DM state, accompanied by the creation
of an electron-hole pair. This perspective is particularly well-suited
for the application of Cutkosky rules~\citep{Cutkosky:1960sp} in
the theoretical analysis. By applying the cutting rules, one can extract
physically meaningful information, such as decay widths or absorption
probabilities, from self-energy corrections. If we consider a DM particle
$\chi$ coupled to an electron via a mediator particle $A'$, the
rate of the initial $\chi$ decays to a lower energy $\chi$ and an
electron-hole pair can be related to the imaginary part of the DM
self-energy diagram. According to the cutting rules, cutting the simplest
two-loop self-energy diagram corresponds to the following leading-order
scattering amplitude: \begin{equation}
\vcenter{\hbox{\tikzset{every picture/.style={line width=0.75pt}}
\begin{tikzpicture}[
  thick,
  DMfermion/.style={draw=gray!70,
    postaction={decorate},
    decoration={
      markings,
      mark=at position 0.57 with {\arrow{Triangle[scale=0.65, reversed]}}
    },
    thick
  },
  fermion/.style={
    postaction={decorate},
    decoration={
      markings,
      mark=at position 0.57 with {\arrow{Triangle[scale=0.65, reversed]}}
    },
    thick
  },
  photon/.style={
    dashed,          
    thick
  },
fermionloop/.style={
    postaction={decorate},
    decoration={
      markings,
      mark=at position 0.82 with {\arrow{Triangle[scale=0.65]}},
      },
    thick
  },
  momentum/.style={
    postaction={decorate},
    decoration={
      markings,
      mark=at position 0.32 with {\arrow{Triangle[scale=0.65]}}
    }
  }
]

\coordinate (A) at (-2,0);
\coordinate (B) at (-1,0);
\coordinate (C) at (1,0);
\coordinate (D) at (2,0);
\coordinate (X) at (0,0.9); 

\draw[DMfermion,line width=1.2pt] (A) node[above] {$$} -- (B) node[midway,below] {};
\draw[DMfermion,line width=1.2pt] (B) -- (C) node[midway,below] {$$} node[right] {};
\draw[DMfermion,line width=1.2pt] (C) -- (D) node[midway,below] {$$} node[right] {};
\filldraw[fill=black] (-1,0) circle[radius=1.2pt];
\filldraw[fill=black] (1,0) circle[radius=1.2pt];

\draw[photon,line width=1.2pt] (B) to[out=90,in=180] coordinate[pos=0.7] (X1) (X);
\draw[photon,line width=1.2pt] (X) to[out=0,in=90] coordinate[pos=0.7] (X2) (C);

\filldraw[white] (X) circle[radius=0.5cm]; 
\draw[fermionloop,line width=1.2pt] (X)  circle[radius=0.5cm];
\path[momentum] (X)  circle[radius=0.5cm];
\draw[gray!70,dotted, line width=1.2pt] 
  ($(X)+(0,-1.4)$) -- ($(X)+(0,1)$); 
\filldraw[fill=black] (-0.49,0.80) circle[radius=1.2pt];
\filldraw[fill=black] (0.49,0.80) circle[radius=1.2pt];
\end{tikzpicture}}}\longrightarrow\vcenter{\hbox{\tikzset{every picture/.style={line width=0.75pt}}
\begin{tikzpicture}[
  thick,
  DMfermion/.style={draw=gray!70,
    postaction={decorate},
    decoration={
      markings,
      mark=at position 0.57 with {\arrow{Triangle[scale=0.65,reversed]}}
    },
    thick
  },
   fermion/.style={
    postaction={decorate},
    decoration={
      markings,
      mark=at position 0.57 with {\arrow{Triangle[scale=0.65, reversed]}}
    },
    thick
  },
  photon/.style={
    dashed,          
    thick
  }]

\coordinate (A) at (-1.1,-0.75);
\coordinate (B) at (-0.65,0);
\coordinate (C) at (-1.1,0.75);
\coordinate (D) at (0.65,0);
\coordinate (E) at (1.1,0.75);
\coordinate (F) at (1.1,-0.75); 

\draw[DMfermion,line width=1.2pt] (B) node[above] {$$} -- (A) node[midway,below] {};
\draw[DMfermion,line width=1.2pt] (C) node[above] {$$} -- (B) node[midway,below] {};
\draw[photon,line width=1.2pt] (B) -- (D) node[midway,below] {$$} node[right] {};
\draw[fermion,line width=1.2pt] (D) -- (E) node[midway,below] {$$} node[right] {};
\draw[fermion,line width=1.2pt] (F) -- (D) node[midway,below] {$$} node[right] {};
\filldraw[fill=black] (B) circle[radius=1.2pt];
\filldraw[fill=black] (D) circle[radius=1.2pt];
\end{tikzpicture}}}\label{eq:cutting0},
\end{equation}where the lines in gray and black represent the DM particle and electron
(hole), respectively, and the mediator particle is denoted with a
dashed line. Eq.~(\ref{eq:cutting0}) indicates that self-energy diagram
on the left corresponds to the specific scattering process on the
right. Therefore, when analyzing a diagram, cutting it can provide
deeper insight into the underlying physical processes. Additionally,
the cutting procedure can be applied to any amplitude, and the generalization
of Eq.~(\ref{eq:cutting0}) provides an efficient shortcut for calculating
the imaginary parts of loop amplitudes.

However, although the cutting procedure in Eq.~(\ref{eq:cutting0})
appears natural, its validity should be justified in condensed matter
systems. There are two significant differences between the case of
relativistic free electrons and that of electrons bound in solids.
First, a realistic description of electronic properties in condensed
matter systems requires wavefunctions that go beyond the simple plane-wave
approximation --- such as those accounting for inhomogeneous electron
distribution in crystal structures. With advances in DFT, the electronic
structures of such systems can now be accurately computed using various
DFT-based approaches. Consequently, the cutting rules must be appropriately
modified to account for these more realistic scenarios. Second, given
that the transverse electromagnetic coupling is negligible for non-relativistic
electrons in solids, and only the Coulomb interaction is taken into
account in DFT, it becomes necessary to establish a self-consistent
set of cutting rules that completely ignore photon emissions.

Moreover, since the cutting rules offer an effective method to prove
the optical theorem on a diagram-by-diagram basis, they can be employed
to demonstrate the unitarity of the interaction between non-relativistic
DM particles and the electron-hole system in solids, even as nontrivial
collective electronic behaviors including screening and plasmon excitations~\citep{Kurinsky:2020dpb,Gelmini_2020,Kozaczuk:2020uzb,Mitridate:2021ctr,Knapen:2021run,Hochberg:2021pkt,Knapen:2021bwg,Liang:2021zkg,Liang:2024xcx}
are taken into consideration, which require the partial resummation
of an infinite series of diagrams and cannot be captured by standard
two-body scattering descriptions.

Encouraged by the motivations outlined above, in this work we presents
a self-consistent framework of cutting rules for electron-hole propagators
derived from DFT-based Kohn--Sham wavefunctions, under the assumption
of Coulomb-only interactions and instantaneous dark matter--electron
couplings. The remainder of this paper is structured as follows. In
Sec.~\ref{sec:DM-electron-interaction-1}, we begin by introducing
the working model of DM-electron interaction that underpins our analysis.
Sec.~\ref{sec:self-energy-and-decay} offers a concise overview of
how to compute the decay rate of an unstable particle from its one-to-one
scattering amplitude. In Sec.~\ref{sec:cutting-rules-scalar}, we
take scalar $\phi^{3}$ theory as a pedagogical example to demonstrate
the derivation of the cutting rules using the largest time equation.
Building on this foundation, Sec.~\ref{sec:-cutting-rules-electron}
extends and establishes the cutting rules for non-relativistic DM
particles and electrons in solid-state systems. We present our summary
and discussions in Sec.~\ref{sec:Conclusions}. Appendix~\ref{sec:appendix1}
collects the relevant Feynman rules used throughout the main text.
Natural units ($\hbar=c=1$) are used throughout the following discussion.

\section{DM-electron interaction\label{sec:DM-electron-interaction-1}}

Our discussion is based on a prototypical DM model, where the DM candidate
is a Dirac particle~($\chi$) coupling to standard model particles
via a massive vector $A_{\mu}^{'}$, with the interaction 
\begin{eqnarray}
\mathcal{L}_{\mathrm{int}} & = & g_{\chi}\bar{\chi}\gamma^{\mu}\chi A_{\mu}^{'}+g_{e}\bar{e}\gamma^{\mu}eA_{\mu}^{'},\label{eq:leptopihilic_interaction}
\end{eqnarray}
where $g_{\chi}$ and $g_{e}$ represent the strengths of the mediator
coupling to the DM particle and electron, respectively. For this generic
type of model, the DM component only couples to the electric density
at the leading order. The effective electron Lagrangian can be obtained
by matching onto the \textit{non-relativistic}~(NR) \textit{effective
field theory}~(EFT) in the beginning~\citep{Mitridate:2021ctr},
which reads as 
\begin{eqnarray}
\mathcal{L}_{A'e}^{\mathrm{NR}} & \supset & g_{e}A_{0}^{'}\psi_{e}^{*}\psi_{e}+\frac{ig_{e}}{2m_{e}}\mathbf{A}^{'}\cdot\left(\psi_{e}^{*}\overrightarrow{\nabla}\psi_{e}-\psi_{e}^{*}\overleftarrow{\nabla}\psi_{e}\right)+\cdots,\label{eq:DM-electron interaction}
\end{eqnarray}
where $\psi_{e}$ is the NR electron wavefunction. In solids, higher-order
terms beyond the leading contribution, such as the second term describing
electric current coupling, are suppressed by the electron velocity
scale $\frac{\nabla}{m_{e}}\sim v_{e}\sim\mathcal{O}\left(10^{-2}\right)$.
Consequently, only the leading-order density-coupling term $g_{e}A_{0}^{'}\psi_{e}^{*}\psi_{e}$
is retained in the effective NR Lagrangian $\mathcal{L}_{A'e}^{\mathrm{NR}}$.
This argument applies equally to halo DM particles, which move in
the NR regime, and thus the DM-electron effective interaction can
be described with a Yukawa potential $V_{\chi e}\left(\mathbf{x}-\mathbf{x}'\right)\sim g_{\chi}g_{e}e^{-m_{A'}\left|\mathbf{x}-\mathbf{x}'\right|}/\left|\mathbf{x}-\mathbf{x}'\right|$.
Likewise, only the longitudinal component in the NR effective electron-photon
interaction
\begin{eqnarray}
\mathcal{L}_{Ae}^{\mathrm{NR}} & \supset & -eA_{0}\psi_{e}^{*}\psi_{e}-\frac{ie}{2m_{e}}\mathbf{A}\cdot\left(\psi_{e}^{*}\overrightarrow{\nabla}\psi_{e}-\psi_{e}^{*}\overleftarrow{\nabla}\psi_{e}\right)+\cdots
\end{eqnarray}
$A_{0}$, which corresponds to the Coulomb interaction, is retained
for the description of electron-electron and electron-ion interaction
in this work. Therefore, in the context of DM direct detection, both
the DM-electron and electron-electron effective interactions are treated
as instantaneous within the non-relativistic solid-state physics framework.

\section{self-energy and decay rate of DM in solids\label{sec:self-energy-and-decay}}

At zero temperature, the theoretical connection between the DM self-energy
and its decay rate in medium is encoded in the following relation:

\begin{equation}
i\mathcal{M}\left(p_{\chi}\rightarrow p_{\chi}\right)\frac{2\pi}{V}\delta_{\mathbf{p}_{\chi},\mathbf{p}_{\chi}'}\delta\left(\varepsilon_{\mathbf{p}_{\chi}}-\varepsilon_{\mathbf{p}_{\chi}'}\right)=
\int\mathrm{d}^{4}z\,\mathrm{d}^{4}z'\left(\sqrt{Z_{\chi}\left(\mathbf{p}_{\chi}\right)}\right)^{2}\left(
\vcenter{\hbox{\tikzset{every picture/.style={line width=0.75pt}}
\begin{tikzpicture}[scale=0.9,
  thick,
  DMfermion/.style={draw=gray!70,
    postaction={decorate},
    decoration={
      markings,
      mark=at position 0.57 with {\arrow{Triangle[scale=0.65, reversed]}}
    },
    thick
  },
  fermion/.style={
    postaction={decorate},
    decoration={
      markings,
      mark=at position 0.57 with {\arrow{Triangle[scale=0.65, reversed]}}
    },
    thick}]

\coordinate (A) at (-2,0);
\coordinate (B) at (-1,0);
\coordinate (C) at (1,0);
\coordinate (D) at (2,0);
\coordinate (X) at (0,0); 

\draw[DMfermion,line width=1.2pt] (A)  -- (B) node[midway,above=2pt,black] {$\mathbf{p}_{\chi}$} node[below left=4pt] {$z$};
\draw[DMfermion,line width=1.2pt] (C)node[below right=2pt] {$z'$} -- (D) node[midway,above=2pt,black] {$\mathbf{p}_{\chi}'$};

\filldraw[lightgray!50] (X)  ellipse (1 and 0.75); 
\draw (X)  ellipse (1 and 0.75);
\node at (X) {$G(z,z')$};
\filldraw[fill=black] (-1,0) circle[radius=1.2pt];
\filldraw[fill=black] (1,0) circle[radius=1.2pt];
\end{tikzpicture}}}
\right),
\label{eq:G_function}
\end{equation}where the DM particle momenta $\mathbf{p}_{\chi}$ and $\mathbf{p}_{\chi}'$
are discretely quantized within the detector volume $V$ under periodic
boundary conditions, $Z_{\chi}$ is the DM field renormalization constant,
\begin{tikzpicture}[baseline={(0,0.05)},scale=0.8,
  thick,
  DMfermion/.style={draw=gray!70,
    postaction={decorate},
    decoration={
      markings,
      mark=at position 0.57 with {\arrow{Triangle[scale=0.65, reversed]}}
    },
    thick
  }]
\coordinate (A) at (-2,0);
\coordinate (B) at (-1,0);
\draw[DMfermion,line width=1.2pt] (A)  -- (B) node[midway,above=2pt,black] {$\mathbf{p}_{\chi}$} node[right=2pt] {$z$};
\filldraw[fill=black] (-1,0) circle[radius=1.2pt];
\end{tikzpicture} and \begin{tikzpicture}[baseline={(0,0.05)},scale=0.8,
  thick,
  DMfermion/.style={draw=gray!70,
    postaction={decorate},
    decoration={
      markings,
      mark=at position 0.57 with {\arrow{Triangle[scale=0.65, reversed]}}
    },
    thick
  }]
\coordinate (C) at (1,0);
\coordinate (D) at (2,0);
\draw[DMfermion,line width=1.2pt] (C)node[left=2pt] {\small$z'$} -- (D) node[midway,above=2pt,black] {\small$\mathbf{p}_{\chi}'$};
\filldraw[fill=black] (1,0) circle[radius=1.2pt];
\end{tikzpicture} represent the DM outgoing external leg $e^{-i\mathbf{p}_{\chi}\cdot\boldsymbol{z}+i\varepsilon_{\mathbf{p}_{\chi}}t_{z}}/\sqrt{V}$
and incoming external leg $e^{i\mathbf{p}'_{\chi}\cdot\boldsymbol{z}'-i\varepsilon_{\mathbf{p}'_{\chi}}t_{z'}}/\sqrt{V}$
, respectively, with $\varepsilon_{\mathbf{p}_{\chi}}=\left|\mathbf{p}_{\chi}\right|^{2}/\left(2m_{\chi}\right)$
($\varepsilon_{\mathbf{p}'_{\chi}}=\left|\mathbf{p}'_{\chi}\right|^{2}/\left(2m_{\chi}\right)$)
being the non-relativistic DM outgoing (incoming) energy, and $\mathcal{M}\left(p_{\chi}\rightarrow p_{\chi}\right)$
is the amplitude for one-to-one DM particle ``scattering''. Note
that the DM-electron coupling strength $g_{e}g_{\chi}$ is so weak
that we take the DM field renormalization constant $Z_{\chi}\simeq1$.
At the leading order of $g_{e}g_{\chi}$, the DM self-energy blob
$G\left(z,z'\right)$ is explicitly expressed as \begin{equation}
\vcenter{\hbox{\tikzset{every picture/.style={line width=0.75pt}}
\begin{tikzpicture}[scale=0.9,
  thick,
  DMfermion/.style={draw=gray!70,
    postaction={decorate},
    decoration={
      markings,
      mark=at position 0.57 with {\arrow{Triangle[scale=0.65, reversed]}}
    },
    thick
  },
  fermion/.style={
    postaction={decorate},
    decoration={
      markings,
      mark=at position 0.57 with {\arrow{Triangle[scale=0.65, reversed]}}
    },
    thick}]

\coordinate (A) at (-2,0);
\coordinate (B) at (-1,0);
\coordinate (C) at (1,0);
\coordinate (D) at (2,0);
\coordinate (X) at (0,0); 


\node at(B)[left=4pt] {$z$};
\node at(C)[right=4pt] {$z'$};

\filldraw[lightgray!50] (X)  ellipse (1 and 0.75); 
\draw (X)  ellipse (1 and 0.75);
\node at (X) {$G(z,z')$};
\filldraw[fill=black] (-1,0) circle[radius=1.2pt];
\filldraw[fill=black] (1,0) circle[radius=1.2pt];
\end{tikzpicture}}}=
\begin{tikzpicture}[baseline={(0,0.0)},scale=0.9,
  thick,
  DMfermion/.style={draw=gray!70,
    postaction={decorate},
    decoration={
      markings,
      mark=at position 0.52 with {\arrow{Triangle[scale=0.65, reversed]}}
    },
    thick
  },
  photon/.style={
    dashed,          
    thick
  }]


\coordinate (B) at (-1,0);
\coordinate (C) at (1,0);
\coordinate (X) at (0,0.8); 

\draw[DMfermion,line width=1.2pt] (B) -- (C) node[midway,below] {$$} node[right] {};
\filldraw[fill=black] (B) circle[radius=1.2pt];
\filldraw[fill=black] (C) circle[radius=1.2pt];

\draw[photon,line width=1.2pt] (B) to[out=90,in=180] coordinate[pos=0.7] (X1) (X);
\draw[photon,line width=1.2pt] (X) to[out=0,in=90] coordinate[pos=0.7] (X2) (C);
\node at(B)[left=4pt] {$z$};
\node at(C)[right=4pt] {$z'$};
\end{tikzpicture}+
\begin{tikzpicture}[baseline={(0,0.0)},scale=0.9,
  thick,
  DMfermion/.style={draw=gray!70,
    postaction={decorate},
    decoration={
      markings,
      mark=at position 0.52 with {\arrow{Triangle[scale=0.65, reversed]}}
    },
    thick
  },
  fermion/.style={
    postaction={decorate},
    decoration={
      markings,
      mark=at position 0.57 with {\arrow{Triangle[scale=0.65, reversed]}}
    },
    thick
  },
  photon/.style={
    dashed,          
    thick
  },
fermionloop/.style={
    postaction={decorate},
    decoration={
      markings},
    thick},
   sparse crosshatch/.style={
    pattern={Hatch[angle=45,distance=4.2pt,line width=0.7pt]},
    pattern color=black
  },
momentum/.style={
    postaction={decorate},
    decoration={
      markings,
      mark=at position 0.42 with {\arrow{Triangle[scale=0.65]}},
      mark=at position 0.68 with {\arrow{Triangle[scale=0.65]}},
    }
  },
momentum1/.style={
    postaction={decorate},
    decoration={
      markings,
      mark=at position 0.12 with {\arrow{Triangle[scale=0.65]}},
      mark=at position 0.88 with {\arrow{Triangle[scale=0.65]}}
    }
  }]

\coordinate (A) at (-3,0);
\coordinate (B) at (-1.5,0);
\coordinate (C) at (1.5,0);
\coordinate (D) at (1.5,0);
\coordinate (X) at (0,1.2); 
\coordinate (aleft) at (-0.5,1.4); 
\coordinate (aright) at (0.5,1.4);
\coordinate (yleft) at (-1,0.85);
\coordinate (yright) at (1,0.85);
\draw[DMfermion,line width=1.2pt] (B) -- (C) node[midway,below] {$$} node[right] {};
\filldraw[fill=black] (B) circle[radius=1.2pt];
\filldraw[fill=black] (C) circle[radius=1.2pt];

\draw[photon,line width=1.2pt] (B) to[out=90,in=180] coordinate[pos=0.7] (X1) (X);
\draw[photon,line width=1.2pt] (X) to[out=0,in=90] coordinate[pos=0.7] (X2) (C);

\filldraw[white] (aleft) arc (90:270:0.5);
\draw (aleft) arc (90:270:0.5);
\filldraw[lightgray!50] (aleft) rectangle (0.5,0.4);
\draw (aleft) rectangle (0.5,0.4);
\filldraw[white] (aright) arc (90:-90:0.5);
\draw (aright) arc (90:-90:0.5);
\path[momentum] (-0.5,0.9)  circle[radius=0.5cm];
\path[momentum1] (0.5,0.9)  circle[radius=0.5cm];
\filldraw[fill=black] (yleft)node[left=5pt] {$y$} circle[radius=1.2pt];
\filldraw[fill=black] (yright) node[right=5pt] {$y'$} circle[radius=1.2pt];
\node at(B)[left=4pt] {$z$};
\node at(C)[right=4pt] {$z'$};
\filldraw[fill=black] (yleft)node[left=5pt] {$$} circle[radius=1.2pt];
\filldraw[fill=black] (yright) node[right=5pt] {$$} circle[radius=1.2pt];
\end{tikzpicture},
\label{eq:G_function1}
\end{equation}where $y$ and $y'$ , as well as other internal vertices in the electron
blob are integrated over. The tadpole diagram contributions associated
with electrons and ions cancel identically for the massive vector
interactions, due to the electron-ion charge neutrality in solids.
The electron blob 
\newcommand{\blobFigure}[1][]{
    \begin{tikzpicture}[
        baseline={(0,0.4)}, 
        scale=1,          
        thick,
        momentum/.style={
            postaction={decorate},
            decoration={
                markings,
                mark=at position 0.42 with {\arrow{Triangle[scale=0.65]}},
                mark=at position 0.68 with {\arrow{Triangle[scale=0.65]}},
            }
        },
        momentum1/.style={
            postaction={decorate},
            decoration={
                markings,
                mark=at position 0.12 with {\arrow{Triangle[scale=0.65]}},
                mark=at position 0.88 with {\arrow{Triangle[scale=0.65]}}
            }
        },
        #1 
    ]

    \coordinate (aleft) at (-0.5,1.4); 
    \coordinate (aright) at (0.5,1.4);
    \coordinate (yleft) at (-1,0.85);
    \coordinate (yright) at (1,0.85);

    \filldraw[white] (aleft) arc (90:270:0.5);
    \draw (aleft) arc (90:270:0.5);

    \filldraw[lightgray!50] (aleft) rectangle (0.5,0.4);
    \draw (aleft) rectangle (0.5,0.4);

    \filldraw[white] (aright) arc (90:-90:0.5);
    \draw (aright) arc (90:-90:0.5);

    \path[momentum] (-0.5,0.9) circle[radius=0.5cm];
    \path[momentum1] (0.5,0.9) circle[radius=0.5cm];

    \filldraw[fill=black] (yleft) node[left=2pt] {\small$$} circle[radius=1.2pt];
    \filldraw[fill=black] (yright) node[right=2pt] {\small$$} circle[radius=1.2pt];

    \end{tikzpicture}%
}
\newcommand{\PI}[2][]{%
  \begin{tikzpicture}[#1] 
    \fill[pattern=north east lines,  pattern color=gray] (0,0) circle (#2);
    \draw[thick] (0,0) circle (#2);
    \coordinate (c1left) at (-0.5,0);
    \coordinate (c1right) at (0.5,0);
    \filldraw[fill=black] (c1left) node[left=2pt] {\small$$} circle[radius=1.2pt];
    \filldraw[fill=black] (c1right) node[right=2pt] {\small$$} circle[radius=1.2pt];
    \end{tikzpicture}%
}
 \begin{equation}
\vcenter{\hbox{\tikzset{every picture/.style={line width=0.75pt}}
\begin{tikzpicture}[
        baseline={(0,0.4)}, scale=0.8,          
        thick,
        momentum/.style={
            postaction={decorate},
            decoration={
                markings,
                mark=at position 0.42 with {\arrow{Triangle[scale=0.65]}},
                mark=at position 0.68 with {\arrow{Triangle[scale=0.65]}},
            }
        },
        momentum1/.style={
            postaction={decorate},
            decoration={
                markings,
                mark=at position 0.12 with {\arrow{Triangle[scale=0.65]}},
                mark=at position 0.88 with {\arrow{Triangle[scale=0.65]}}
            }
        }]

    \coordinate (aleft) at (-0.5,1.4); 
    \coordinate (aright) at (0.5,1.4);
    \coordinate (yleft) at (-1,0.85);
    \coordinate (yright) at (1,0.85);

    \filldraw[white] (aleft) arc (90:270:0.5);
    \draw (aleft) arc (90:270:0.5);

    \filldraw[lightgray!50] (aleft) rectangle (0.5,0.4);
    \draw (aleft) rectangle (0.5,0.4);

    \filldraw[white] (aright) arc (90:-90:0.5);
    \draw (aright) arc (90:-90:0.5);

    \path[momentum] (-0.5,0.9) circle[radius=0.5cm];
    \path[momentum1] (0.5,0.9) circle[radius=0.5cm];

    \filldraw[fill=black] (yleft) node[left=2pt] {\small$$} circle[radius=1.2pt];
    \filldraw[fill=black] (yright) node[right=2pt] {\small$$} circle[radius=1.2pt];

    \end{tikzpicture}}}=
\vcenter{\hbox{\tikzset{
    vertex/.pic={
        \fill[pattern=north east lines, pattern color=gray] (0,0) circle (0.4);
        \draw[thick] (0,0) circle (0.4);
        \filldraw[fill=black] (-0.4,0) circle[radius=1.2pt];
        \filldraw[fill=black] (0.4,0) circle[radius=1.2pt];
    },
    photon/.style={
        decorate, decoration={snake, amplitude=2pt, segment length=4pt},
        line width=1.2pt 
    }
}
\begin{tikzpicture}
    \pic at (0,0) {vertex};
    \end{tikzpicture}
}}+
\vcenter{\hbox{\tikzset{
    vertex/.pic={
        \fill[pattern=north east lines, pattern color=gray] (0,0) circle (0.4);
        \draw[thick] (0,0) circle (0.4);
        \filldraw[fill=black] (-0.4,0) circle[radius=1.2pt];
        \filldraw[fill=black] (0.4,0) circle[radius=1.2pt];
    },
    photon/.style={
        decorate, decoration={snake, amplitude=2pt, segment length=4pt},
        line width=1.2pt 
    }
}
\begin{tikzpicture}
    \pic at (0,0) {vertex};
    \draw[photon] (0.4,0) -- (1.2,0);
    \pic at (1.6,0) {vertex};
    \end{tikzpicture}
}}+
\vcenter{\hbox{\tikzset{
    vertex/.pic={
        \fill[pattern=north east lines, pattern color=gray] (0,0) circle (0.4);
        \draw[thick] (0,0) circle (0.4);
        \filldraw[fill=black] (-0.4,0) circle[radius=1.2pt];
        \filldraw[fill=black] (0.4,0) circle[radius=1.2pt];
    },
    photon/.style={
        decorate, decoration={snake, amplitude=2pt, segment length=4pt},
        line width=1.2pt 
    }
}
\begin{tikzpicture}
    \pic at (0,0) {vertex};
    \draw[photon] (0.4,0) -- (1.2,0);
    \pic at (1.6,0) {vertex};
    \draw[photon] (2.0,0) -- (2.8,0);
    \pic at (3.2,0) {vertex};
\end{tikzpicture}
}}+\cdots \label{eq:blob_series}
\end{equation} represents the sum of all possible diagrams between the vertices
$y$ and $y'$, which can be summarized as a sum of a series of one-particle-irreducible
(1PI)  \PI[baseline=-0.5ex,scale=0.5]{0.5}, with \begin{tikzpicture}[baseline={(0,-0.1)},scale=0.8,
 photon/.style={
        decorate, decoration={snake, amplitude=2pt, segment length=4pt},
        line width=1.2pt 
    }]
\draw[photon] (-0.5,0) -- (0.5,0);
\end{tikzpicture}
 being the Coulomb interaction
\begin{eqnarray}
D^{\mathrm{Cou}}\left(x,y\right) & = & \int\sum_{\mathbf{q}}\frac{i}{\left|\mathbf{q}\right|^{2}}\frac{e^{i\mathbf{q}\cdot\left(\mathbf{x}-\mathbf{y}\right)}}{V}\frac{e^{-i\omega\left(t_{x}-t_{y}\right)}\mathrm{d}\omega}{2\pi},
\end{eqnarray}
and \begin{tikzpicture}[baseline={(0,-0.1)},scale=0.8,
 photon/.style={
        dashed,          
        line width=1.2pt 
    }]
\draw[photon] (-0.5,0) -- (0.5,0);
\end{tikzpicture}
being the massive mediator propagator 
\begin{eqnarray}
D^{A'}\left(x,y\right) & = & \int\sum_{\mathbf{q}}\frac{-i}{\left|\mathbf{q}\right|^{2}+m_{A'}^{2}}\frac{e^{i\mathbf{q}\cdot\left(\mathbf{x}-\mathbf{y}\right)}}{V}\frac{e^{-i\omega\left(t_{x}-t_{y}\right)}\mathrm{d}\omega}{2\pi}.
\end{eqnarray}
If the spin is neglected, the electron-hole propagator in solids \begin{tikzpicture}[baseline={(0,-0.1)}, scale=1]
\draw[line width=1.0pt,
    decoration={markings, mark=at position 0.57 with {\arrow{Triangle[scale=0.65]}}},
    postaction={decorate}
  ] 
  (0.5,0) -- (-0.5,0) ;
  \filldraw[black] (-0.5,0) circle[radius=1.2pt]
    node[left=2pt] {$x$};
  \filldraw[black] (0.5,0) circle[radius=1.2pt]
    node[right=2pt] {$y$};
\end{tikzpicture} is given as
\begin{eqnarray}
D_{e}\left(x,y\right) & = & \int\sum\limits_{k}\frac{iu_{k}\left(\mathbf{x}\right)u_{k}^{*}\left(\mathbf{y}\right)}{\omega-\varepsilon_{k}+i\eta_{k}\,0^{+}}\frac{e^{-i\omega\left(t_{x}-t_{y}\right)}\mathrm{d}\omega}{2\pi},\label{eq:e-h_propagator}
\end{eqnarray}
with $u_{k}\left(\mathbf{x}\right)$ and $\varepsilon_{k}$ representing
the $k$-th normalized \emph{Kohn-Sham} (KS) wavefunction and energy
eigenvalue of the electron-hole system, respectively, which satisfies
the KS one-particle equation
\begin{eqnarray}
\left[-\frac{\nabla^{2}}{2m_{e}}+v_{\mathrm{ext}}\left(\mathbf{x}\right)+v_{\mathrm{H}}\left(\mathbf{x}\right)+v_{\mathrm{xc}}\left(\mathbf{x}\right)\right]u_{k}\left(\mathbf{x}\right) & = & \varepsilon_{k}u_{k}\left(\mathbf{x}\right),
\end{eqnarray}
where $v_{\mathrm{ext}}\left(\mathbf{x}\right)$ is the external potential
imposed by the ions, $v_{\mathrm{H}}\left(\mathbf{x}\right)=\alpha\sum_{k\leq F}\int\mathrm{d}^{3}x'\left|u_{k}\left(\mathbf{x}'\right)\right|^{2}/\left|\mathbf{x}-\mathbf{x}'\right|$
is the Hartree potential (with $\alpha=e^{2}/\left(4\pi\right)$ being
the electromagnetic fine structure constant, and the sum runs over
all occupied orbitals), and $v_{\mathrm{xc}}\left(\mathbf{x}\right)$
denotes the the local exchange--correlation potential; $\eta_{k}\equiv\mathrm{sgn}\left(k-F\right)$,
which depends on whether the $k$-th state is above or below the Fermi
surface. Given the propagator in Eq.~(\ref{eq:e-h_propagator}),
the Lagrangian for the electron-hole system can be recast as
\begin{eqnarray}
\mathcal{L}_{e} & = & \psi_{e}^{*}\left(x\right)\left[-\frac{\nabla^{2}}{2m_{e}}+v_{\mathrm{ext}}\left(\mathbf{x}\right)+v_{\mathrm{H}}\left(\mathbf{x}\right)+v_{\mathrm{xc}}\left(\mathbf{x}\right)\right]\psi_{e}\left(x\right)-\psi_{e}^{*}\left(\mathbf{x}\right)\left[v_{\mathrm{H}}\left(\mathbf{x}\right)+v_{\mathrm{xc}}\left(\mathbf{x}\right)\right]\psi_{e}\left(\mathbf{x}\right)\nonumber \\
\nonumber \\ &  & -\frac{1}{2}\int\mathrm{d}^{3}x'\psi_{e}^{*}\left(x\right)\psi_{e}\left(x\right)V_{e}\left(\mathbf{x}-\mathbf{x}'\right)\psi_{e}^{*}\left(x'\right)\psi_{e}\left(x'\right),
\end{eqnarray}
where $V_{e}\left(\mathbf{\mathbf{x}-\mathbf{x}'}\right)=\alpha/\left|\mathbf{x}-\mathbf{x}'\right|$
represents the electron-electron Coulomb interaction. In the context
of perturbation theory, $-\psi_{e}^{*}\left(\mathbf{x}\right)\left[v_{\mathrm{H}}\left(\mathbf{x}\right)-v_{\mathrm{xc}}\left(\mathbf{x}\right)\right]\psi_{e}\left(\mathbf{x}\right)$
serves as a perturbative external source term that is relevant for
the calculation of self-energy, and hence the quasi-particle energy.
For example, the Hartree-Fock (HF) self-energy yields \begin{equation}
\vcenter{\hbox{\tikzset{every picture/.style={line width=0.75pt}}
\begin{tikzpicture}[scale=1.2,
  thick,
  DMfermion/.style={draw=gray!70,
    postaction={decorate},
    decoration={
      markings,
      mark=at position 0.57 with {\arrow{Triangle[scale=0.65, reversed]}}
    },
    thick
  },
  fermion/.style={
    postaction={decorate},
    decoration={
      markings,
      mark=at position 0.57 with {\arrow{Triangle[scale=0.65, reversed]}}
    },
    thick}]

\coordinate (X) at (0,0); 

\filldraw[lightgray!50] (X)  ellipse (0.6 and 0.45); 
\draw (X)  ellipse (0.6 and 0.45);
\node at (X) {$\Sigma$};
\filldraw[fill=black] (-0.6,0) circle[radius=1.0pt];
\filldraw[fill=black] (0.6,0) circle[radius=1.0pt];
\end{tikzpicture}}}~=~
~
\begin{tikzpicture}[baseline={(0,0)},scale=1.2,
  thick,
  fermion/.style={
    postaction={decorate},
    decoration={
      markings,
      mark=at position 0.52 with {\arrow{Triangle[scale=0.65, reversed]}}
    },
    thick
  }]


\coordinate (B) at (-0.6,0);
\coordinate (C) at (0.6,0);
\coordinate (X) at (0,0.8); 

\draw[fermion,line width=1.2pt] (B) -- (C) node[midway,below] {$$} node[right] {};
\filldraw[fill=black] (B) circle[radius=1.0pt];
\filldraw[fill=black] (C) circle[radius=1.0pt];
\draw[
    decoration={snake, amplitude=0.8mm, segment length=2.0mm}, 
    decorate,
    line width=1.2pt
  ]
  (-0.6,0) to[out=75, in=105] (0.6,0);
\end{tikzpicture}~-~
\begin{tikzpicture}[baseline={(0,-0.1)},scale=1.2,
  thick,
  fermion/.style={
    postaction={decorate},
    decoration={
      markings,
      mark=at position 0.52 with {\arrow{Triangle[scale=0.65, reversed]}}
    },
    thick
  }]


\coordinate (B) at (-0.6,0);
\coordinate (C) at (0.6,0);
\coordinate (D) at (0,0);

\filldraw[white] (D) circle[radius=0.1cm]; 
\draw[line width=1.0pt] (0,0) circle (0.1);
\draw[line width=1.0pt] (-0.0707, 0.0707) -- (0.0707, -0.0707); 
\draw[line width=1.0pt] (-0.0707, -0.0707) -- (0.0707, 0.0707);
\end{tikzpicture}~,
\label{eq:Sigma}
\end{equation}
where ~$\vcenter{\hbox{\tikzset{every picture/.style={line width=0.75pt}}
\begin{tikzpicture}[scale=0.6,
  thick,
  DMfermion/.style={draw=gray!70,
    postaction={decorate},
    decoration={
      markings,
      mark=at position 0.57 with {\arrow{Triangle[scale=0.65, reversed]}}
    },
    thick
  },
  fermion/.style={
    postaction={decorate},
    decoration={
      markings,
      mark=at position 0.57 with {\arrow{Triangle[scale=0.65, reversed]}}
    },
    thick}]

\coordinate (X) at (0,0); 
\filldraw[lightgray!50] (X)  ellipse (0.75 and 0.55); 
\draw (X)  ellipse (0.75 and 0.55);
\node at (X) {\small$\Sigma$};
\filldraw[fill=black] (-0.75,0) circle[radius=1.2pt];
\filldraw[fill=black] (0.75,0) circle[radius=1.2pt];
\end{tikzpicture}}}$~ represents the electron 1PI, ~\begin{tikzpicture}[baseline={(0,-0.1)},scale=1,
  thick]
\coordinate (D) at (0,0);
\draw[line width=1.0pt] (0,0) circle (0.1);
\draw[line width=1.0pt] (-0.0707, 0.0707) -- (0.0707, -0.0707); 
\draw[line width=1.0pt] (-0.0707, -0.0707) -- (0.0707, 0.0707);
\end{tikzpicture}~ denotes exchange-correlation potential $v_{\mathrm{xc}}$ in DFT,
and the Hartree potential $v_{\mathrm{H}}$ is canceled by the tadpole
diagram \begin{tikzpicture}[baseline={(0,0)},scale=0.6,
  thick,
  fermion/.style={
    postaction={decorate},
    decoration={
      markings,
      mark=at position 0.52 with {\arrow{Triangle[scale=0.65, reversed]}}
    },
    thick
  },
  fermionloop/.style={
    postaction={decorate},
    decoration={
      markings},
    thick},
  photon/.style={
    dashed,          
    thick
  },
  momentum/.style={
    postaction={decorate},
    decoration={
      markings,
      mark=at position 0.3 with {\arrow{Triangle[scale=0.65]}},
      }
  }]


\coordinate (D) at (0,0);
\coordinate (E) at (0,0.4);

\filldraw[fill=black] (D) circle[radius=1.2pt];
\filldraw[fill=black] (E) circle[radius=1.2pt];
\draw[
    decoration={snake, amplitude=0.6mm, segment length=1.2mm}, 
    decorate,
    line width=1.2pt
  ]
  (D) to (0,0.4);
\coordinate (X) at (0,0.7); 
\draw[fermionloop,line width=1.2pt] (X)  circle[radius=0.3cm];
\path[momentum] (X)  circle[radius=0.3cm];
\end{tikzpicture}. In the one-shot perturbative framework, the above external term is
taken into consideration to obtain the electron-hole quasi-particle
energies $\left\{ \tilde{\varepsilon}_{k}\right\} $, which can be
determined in a non-self-consistent way as follows, 
\begin{eqnarray}
\tilde{\varepsilon}_{k} & = & \varepsilon_{k}+\mathrm{Re}\left.\Sigma\left(\omega\right)\right|_{\omega=\varepsilon_{k}},
\end{eqnarray}
where electron-hole quasi-particle self-energy $\Sigma$ is related
to the one-to-one scattering amplitude for the $k$-th KS orbital
through $\mathcal{M}\left(k\rightarrow k\right)=-Z_{e}\left.\Sigma\left(\omega\right)\right|_{\omega=\varepsilon_{k}}$,
and the electron field renormalization constant can be explicitly
approximated as $Z_{e}\left(\varepsilon_{k}\right)=\left(1-\left.\mathrm{d}\Sigma/\mathrm{d}\omega\right|_{\omega=\varepsilon_{k}}\right)^{-1}$
using the first-order Taylor expansion. Beyond the HF self-energy,
one can replace the bare Coulomb potential in Eq.~(\ref{eq:Sigma})
with a corrected Coulomb potential so as to describe electron-hole
quasi-particles with a finite lifetime, and thus the decay rate for
the electron-hole quasi-particle corresponding to the $k$-th KS orbital
can be expressed as the imaginary part of the one-to-one particle
scattering amplitude $\Gamma_{k}=2\,\mathrm{Im}\left[\mathcal{M}\left(k\rightarrow k\right)\right]/V$.

Similarly, the decay rate for a DM with four-momentum $p_{\chi}$
in medium can be written as 
\begin{eqnarray}
\Gamma_{p_{\chi}} & = & \frac{2}{V}\,\mathrm{Im}\left[\mathcal{M}\left(p_{\chi}\rightarrow p_{\chi}\right)\right].\label{eq:decay_rate}
\end{eqnarray}
In subsequent sections, we will explore how to establish a framework
to calculate this DM decay rate based on the cutting rules. 

\section{cutting rules for relativistic scalar fields\label{sec:cutting-rules-scalar}}

Before constructing a set of cutting rules for non-relativistic DM
particles and electrons in solids, in this section we first take a
review on the ``largest time equation'' method introduced by Veltman~\citep{Veltman:1963th}.
We then show how to prove the cutting rules for a relativistic scalar
field $\phi$ with interaction $\mathcal{L}_{\mathrm{int}}=-\lambda\phi^{3}/3!$\footnote{More detailed discussion can be found in  Refs.~(\citep{Veltman:1994wz,Bellac:2011kqa,Nastase_2019})},
by use of this method. 

One first recalls that the Feynman propagator for the field $\phi$
can be written as 
\begin{eqnarray}
D\left(x-y\right) & = & \Theta\left(x^{0}-y^{0}\right)D^{>}\left(x,y\right)+\Theta\left(y^{0}-x^{0}\right)D^{<}\left(x,y\right),
\end{eqnarray}
where $\Theta$ is the Heaviside step function, 
\begin{eqnarray}
D^{>}\left(x,y\right) & = & \int\frac{\mathrm{d}^{3}p}{\left(2\pi\right)^{3}2E_{\mathbf{p}}}e^{-ip\left(x-y\right)},
\end{eqnarray}
and 
\begin{eqnarray}
D^{<}\left(x,y\right) & = & \int\frac{\mathrm{d}^{3}p}{\left(2\pi\right)^{3}2E_{\mathbf{p}}}e^{ip\left(x-y\right)},
\end{eqnarray}
with $p^{0}=E_{\mathbf{p}}=\sqrt{m_{\phi}^{2}+\left|\mathbf{p}\right|^{2}}$
being the energy of the scalar particle.

Now consider a diagram in position space with $N$ points $x_{1}$,
$x_{2}$, $\cdots,$ $x_{N}$, which can be described with an integrand
$G\left(x_{1},x_{2},\cdots,x_{N}\right)$ and is composed of Feynman
propagators $D\left(x_{i}-x_{j}\right)$ connecting points $i$ and
$j$. In order to obtain the cutting rules, the underlining operation
on the vertices in $G\left(x_{1},x_{2},\cdots,x_{N}\right)$ is introduced:
\begin{enumerate}
\item If the vertices $x_{i}$ and $x_{j}$ connected by the propagator
$D\left(x_{i},x_{j}\right)$ are not underlined, one keeps the original
propagator $D\left(x_{i},x_{j}\right)$.
\item If $x_{i}$ is underlined while $x_{j}$ is not underlined, the two
vertices are linked by $D^{>}\left(x_{i},x_{j}\right).$ 
\item If $x_{j}$ is underlined while $x_{i}$ is not underlined, the two
vertices are linked by $D^{<}\left(x_{i},x_{j}\right).$
\item If both $x_{i}$ and $x_{j}$ are underlined, one uses the complex
conjugate $D\left(x_{i},x_{j}\right)^{*}$ of the propagator.
\item If $x_{i}$ is underlined, the factor $-i\lambda$ associated with
the original vertex is changed to its complex conjugate $i\lambda$. 
\end{enumerate}
Rules (2-4) can be summarized as follows, \begin{subequations}
\begin{align}
D\left(\underline{x}_{i},x_{j}\right)&=D^{>}\left(x_{i},x_{j}\right)\\
D\left(x_{i},\underline{x}_{j}\right)&=D^{<}\left(x_{i},x_{j}\right)\\
D\left(\underline{x}_{i},\underline{x}_{j}\right)&=D\left(x_{i},x_{j}\right)^{*}.\label{eq:scalar*}
\end{align}
\end{subequations}These rules leads to the so-called ``largest time equation'' if
one assumes that the time coordinate $x_{i}^{0}$ is larger than any
other time coordinate $x_{j}^{0}$: \begin{subequations}
\begin{align}
D\left(\underline{x}_{i},x_{j}\right)&=D^{>}\left(x_{i},x_{j}\right)=D\left(x_{i},x_{j}\right),\label{eq:Ltime1} \\
D\left(x_{j},\underline{x}_{i}\right)&=D^{<}\left(x_{j},x_{i}\right)=D\left(x_{j},x_{i}\right),\\
D\left(\underline{x}_{i},\underline{x}_{j}\right)&=D\left(x_{i},x_{j}\right)^{*}=D^{<}\left(x_{i},x_{j}\right)=D\left(x_{i},\underline{x}_{j}\right),\\
D\left(\underline{x}_{j},\underline{x}_{i}\right)&=D\left(x_{j},x_{i}\right)^{*}=D^{>}\left(x_{j},x_{i}\right)=D\left(\underline{x}_{j},x_{i}\right),\label{eq:Ltime2}
\end{align}
\end{subequations}where the relation $D^{>}\left(x,y\right)^{*}=D^{<}\left(x,y\right)$
is used. Combining the largest time equation and the above rule (5)
it is evident that for the integrand $G$ with an arbitrary set of
$\left\{ x_{k}\right\} $ underlined (say, $G\left(\underline{x}_{1},x_{2},\cdots,x_{i},\cdots,x_{N}\right)$
for example), one always has $G\left(\underline{x}_{1},x_{2},\cdots,\underline{x}_{i},\cdots,x_{N}\right)=-G\left(\underline{x}_{1},x_{2},\cdots,x_{i},\cdots,x_{N}\right)$,
provided that $x_{i}$, whose time component $x_{i}^{0}$ is the largest
among all coordinates, is underlined. Then we have
\begin{eqnarray}
\sum G\left(x_{1},\cdots,\underline{x}_{k},\cdots,x_{N}\right) & = & 0,\label{eq:sum_zero-1}
\end{eqnarray}
where the sum is over all possible underlining configurations. Then
we consider the particular configuration where all vertices are underlined,
\emph{i.e}., $G\left(\underline{x}_{1},\cdots,\underline{x}_{k},\cdots,\underline{x}_{N}\right)$.
By using the underlining rule (5) and Eq.~(\ref{eq:scalar*}), it
is straightforward to see that $G\left(\underline{x}_{1},\cdots,\underline{x}_{k},\cdots,\underline{x}_{N}\right)=G\left(x_{1},\cdots,x_{k},\cdots,x_{N}\right)^{*}$.
Thus Eq.~(\ref{eq:sum_zero-1}) can be rearranged as 

\begin{eqnarray}
G\left(x_{1},\cdots,x_{N}\right)+G\left(x_{1},\cdots,x_{N}\right)^{*} & = & -\sum_{\mathrm{partial}}G\left(x_{1},\cdots,\underline{x}_{k},\cdots,x_{N}\right),\label{eq:sum_zero-2}
\end{eqnarray}
where on the RHS, the sum is taken over all possibilities in which
both underlined and non-underlined vertices are included in each configuration---in
other words, the terms on the RHS are all partially underlined.

In order to obtain Feynman diagrams in momentum space and scattering
amplitudes, we multiply the integrand $G\left(x_{1},\cdots,x_{N}\right)$
by the factors $\left\{ e^{-ip_{n}x_{n}}\right\} $ ($\left\{ e^{ik_{m}x_{m}}\right\} $)
for four-momenta $\left\{ p_{n}\right\} $ ($\left\{ k_{m}\right\} $)
flowing into (out of) the diagram, and integrate over all vertices
$\left(x_{1},\cdots,x_{N}\right)$, \emph{i.e.},
\begin{eqnarray}
\left(2\pi\right)^{4}\delta^{4}\left(\sum_{n}p_{n}-\sum_{m}k_{m}\right)i\mathcal{M}\left(\left\{ p_{n}\right\} \rightarrow\left\{ k_{m}\right\} \right) & = & \int\prod_{i=1}^{N}\mathrm{d}^{4}x_{i}\,G\left(x_{1},\cdots,x_{N}\right)e^{-i\left(\sum_{n}p_{n}x_{n}-\sum_{m}k_{m}x_{m}\right)}.\nonumber \\
\end{eqnarray}
 Applying the same operation to the second term on the LHS of Eq.~(\ref{eq:sum_zero-2})
gives 
\begin{align}
\int\prod_{i=1}^{N}\mathrm{d}^{4}x_{i}G^{*}\left(x_{1},\cdots,x_{N}\right)e^{-i\left(\sum_{n}p_{n}x_{n}-\sum_{m}k_{m}x_{m}\right)}= & \int\prod_{i=1}^{N}\mathrm{d}^{4}x_{i}\left[G\left(x_{1},\cdots,x_{N}\right)e^{i\left(\sum_{n}p_{n}x_{n}-\sum_{m}k_{m}x_{m}\right)}\right]^{*}\nonumber \\
=-\left(2\pi\right)^{4}\delta^{4}\left(\sum_{n}p_{n}-\sum_{m}k_{m}\right)i\mathcal{M}^{*}\left(\left\{ k_{m}\right\} \rightarrow\left\{ p_{n}\right\} \right),
\end{align}
and thus Eq.~(\ref{eq:sum_zero-2}) in momentum space reads as 
\begin{align}
 & \left(2\pi\right)^{4}\delta^{4}\left(\sum_{n}p_{n}-\sum_{m}k_{m}\right)\left[i\mathcal{M}\left(\left\{ p_{n}\right\} \rightarrow\left\{ k_{m}\right\} \right)-i\mathcal{M}^{*}\left(\left\{ k_{m}\right\} \rightarrow\left\{ p_{n}\right\} \right)\right]\nonumber \\
= & -\sum_{\mathrm{partial}}\int\prod_{i=1}^{N}\mathrm{d}^{4}x_{i}\,G\left(x_{1},\cdots,\underline{x}_{k},\cdots,x_{N}\right)e^{-i\left(\sum_{n}p_{n}x_{n}-\sum_{m}k_{m}x_{m}\right)},\label{eq: optical_cutting_rules}
\end{align}
where the LHS is recognized as $i\mathcal{M}\left(a\rightarrow b\right)-i\mathcal{M}^{*}\left(b\rightarrow a\right)$
times an overall delta function $\left(2\pi\right)^{4}$ $\times\delta^{4}\left(\sum_{n}p_{n}-\sum_{m}k_{m}\right)$.
Comparing this equation with the optical theorem
\begin{eqnarray}
i\mathcal{M}\left(a\rightarrow b\right)-i\mathcal{M}^{*}\left(b\rightarrow a\right) & = & -\sum_{f}\int\mathrm{d}\Pi_{f}\mathcal{M}^{*}\left(b\rightarrow f\right)\mathcal{M}\left(a\rightarrow f\right),\label{eq:optial theorem}
\end{eqnarray}
where the sum runs over all possible sets $f$ of final-state particles,
and $\mathrm{d}\Pi_{f}$ denotes the relevant differential final-state
phase space, it is observed that if the optical theorem holds for
any particular diagram, the RHS of Eq.~(\ref{eq: optical_cutting_rules})
should give the RHS of Eq.~(\ref{eq:optial theorem}). 
\begin{figure}
\begin{centering}
\includegraphics[scale=0.35]{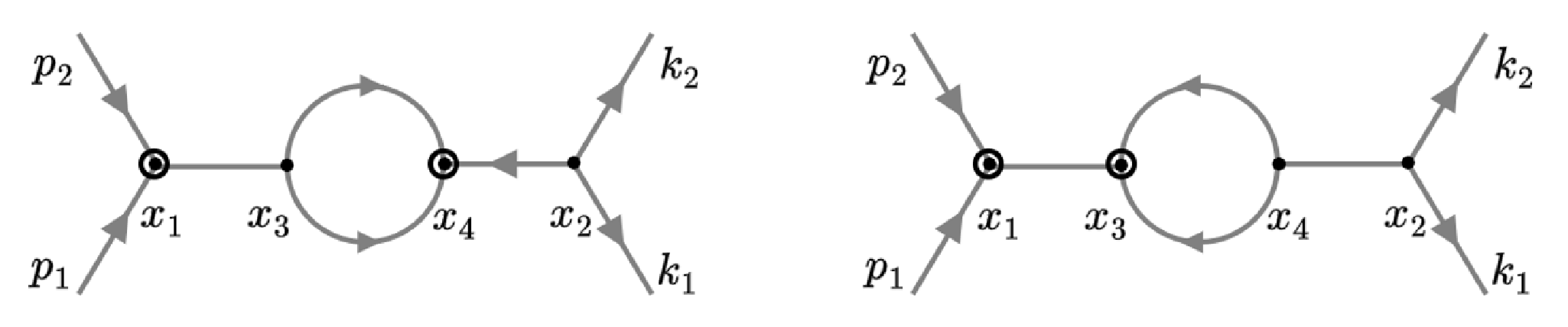}
\par\end{centering}
\caption{\label{fig:scalar_circled}Two specific ways of underlining vertices
in a diagram for the scattering process $p_{1}p_{2}\rightarrow k_{1}k_{2}$.}

\end{figure}

In fact, the underlined propagators in $G\left(x_{1},\cdots,\underline{x}_{k},\cdots,x_{N}\right)$
can significantly simplify the calculation of Eq.~(\ref{eq: optical_cutting_rules})
through the argument of energy conservation. We take the diagrams
in Fig.~\ref{fig:scalar_circled} as examples to illustrate this
point. In these diagrams, circular annotations are used to indicate
vertices underlined in Eq.~(\ref{eq: optical_cutting_rules}). Accordingly,
the left diagram corresponds to the term $\int\prod_{i=1}^{4}\mathrm{d}^{4}x_{i}\,G\left(\underline{x}_{1},x_{2},x_{3},\underline{x}_{4}\right)e^{-i\left[\left(p_{1}+p_{2}\right)x_{1}-\left(k_{1}+k_{2}\right)x_{2}\right]}$,
while the right diagram represents the term $\int\prod_{i=1}^{4}\mathrm{d}^{4}x_{i}\,G\left(\underline{x}_{1},x_{2},\underline{x}_{3},x_{4}\right)e^{-i\left[\left(p_{1}+p_{2}\right)x_{1}-\left(k_{1}+k_{2}\right)x_{2}\right]}$.
In addition, following the convention that $e^{-ipx}$ denotes momentum
$p$ flowing into vertex $x$, we use arrows in diagrams to indicate
momentum flow through underlined scalar propagators and external legs.
Thus, the left diagram shows a net momentum flowing out of vertex
$x_{2}$, violating the energy conservation. In other words, the integration
over $x_{2}$ yields a delta $\delta\left(p^{0}+k_{1}^{0}+k_{2}^{0}\right)=0$
because all energies $\left\{ p^{0},\,k_{1}^{0},\,k_{2}^{0}\right\} $
are positive definite. Then one can concludes that the term corresponding
to the left diagram vanishes. On the other hand, the right diagram
shows a net momentum flowing out of vertices $x_{2}$ and $x_{4}$,
so the corresponding term also vanishes by the same reasoning. The
same argument applies to all remaining diagrams. As a result, only
the term corresponding to the diagram where vertices $x_{2}$ and
$x_{4}$ are underlined survives, which, by the underlining rules,
gives \begin{align}
& -\int\prod_{i=1}^{N}\mathrm{d}^{4}x_{i}\,G\left(x_{1},\underline{x}_{2},x_{3},\underline{x}_{4}\right)e^{-i\left[\left(p_{1}+p_{2}\right)x_{1}-\left(k_{1}+k_{2}\right)x_{2}\right]} \nonumber \\[10pt]
={} & -\int\prod_{i=1}^{4}\mathrm{d}^{4}x_{i}\left(
\begin{tikzpicture}[baseline={(0,0)},scale=0.9,
  thick,
  fermion/.style={
    postaction={decorate},
    decoration={
      markings,
      mark=at position 0.57 with {\arrow{Triangle[scale=0.65, reversed]}}
    },
    thick,color=gray!100
  },
  scalar/.style={
    thick,color=gray!100
  },
fermionloop/.style={
    postaction={decorate},
    decoration={
      markings},
   thick,color=gray!100},
momentum/.style={
    postaction={decorate},
    decoration={
      markings,
      mark=at position 0.27 with {\arrow{Triangle[scale=0.85, reversed]}},
      mark=at position 0.77 with {\arrow{Triangle[scale=0.85]}}
    },
      color=gray!100
  },
 photon/.style={
  thick,
  decorate,                  
  decoration={
    snake,                   
    amplitude=1.5pt,           
    segment length=5pt       
  }}]
\coordinate (A) at (-1.5,0);
\coordinate (B) at (1.5,0);
\coordinate (X) at (0,0); 
\coordinate (p1) at (-2,-0.8);
\coordinate (p2) at (-2,0.8);
\coordinate (k1) at (2,-0.8);
\coordinate (k2) at (2,0.8);
\draw[scalar,line width=1.2pt] (A) --(-0.5,0);
\draw[scalar,line width=1.2pt] (0.5,0) --(B);
\draw[fermion,line width=1.2pt] (A)--(p1)node[anchor=center, xshift=-10pt, yshift=3pt, text=black] {$p_{1}$};
\draw[fermion,line width=1.2pt] (A) --(p2)node[anchor=center, xshift=-10pt, yshift=-3pt, text=black] {$p_{2}$};
\draw[fermion,line width=1.2pt] (k1)node[anchor=center, xshift=10pt, yshift=3pt, text=black] {$k_{1}$} --(B);
\draw[fermion,line width=1.2pt] (k2)node[anchor=center, xshift=10pt, yshift=-3pt, text=black] {$k_{2}$} --(B);
\filldraw[white] (X) circle[radius=0.5cm]; 
\draw[fermionloop,line width=1.2pt] (X)  circle[radius=0.5cm];
\path[momentum] (X)  circle[radius=0.5cm];
\filldraw[fill=black] (-0.5,0) node[anchor=center, xshift=-3pt, yshift=-12pt] {$x_{3}$} circle[radius=1.2pt];
\filldraw[fill=black] (0.5,0) node[anchor=center, xshift=3pt, yshift=-12pt] {$x_{4}$} circle[radius=1.2pt];
\filldraw[fill=black] (A) node[anchor=center, xshift=3pt, yshift=-12pt] {$x_{1}$} circle[radius=1.2pt];
\filldraw[fill=black] (B) node[anchor=center, xshift=-3pt, yshift=-12pt] {$x_{2}$} circle[radius=1.2pt];
\draw[line width=1.2pt] (0.5,0) circle[radius=0.12];
\draw[line width=1.2pt] (B) circle[radius=0.12];
\end{tikzpicture}\right) \nonumber \\[10pt]
={} & -\left(\prod_{i=1}^{2}\int\frac{\mathrm{d}^{3}q_{i}}{\left(2\pi\right)^{3}}\frac{1}{2E_{\mathbf{q}_{i}}}\right)\left(2\pi\right)^{4}\delta^{4}\left(p_{1}+p_{2}-q_{1}-q_{2}\right)\mathcal{M}\left(p_{1}p_{2}\rightarrow q_{1}q_{2}\right)\mathcal{M}^{*}\left(k_{1}k_{2}\rightarrow q_{1}q_{2}\right)\nonumber \\
& \times\left(2\pi\right)^{4}\delta^{4}\left(p_{1}+p_{2}-k_{1}-k_{2}\right),
\end{align}
 an expression in the same form of the RHS of Eq.~(\ref{eq:optial theorem}). 

Therefore, by applying the underlining rules and the largest time
equation to this simple example, we have demonstrated unitarity for
scalars, as well as established the cutting rules: one can draw a
cut through all internal lines that carry arrows, but not through
those without; only diagrams are relevant in which on one side of
the cut all vertices are non-circled, and on the other side, all are
circled with external energy flowing from the non-circled to the circled
side.

\section{cutting rules for DM and electrons in solids\label{sec:-cutting-rules-electron}}

In the following, we propose to extend the cutting rules derived for
relativistic scalar fields to non-relativistic DM particles and electron-hole
quasi-particles in solid-state systems. The electron-hole propagator
in solids in Eq.~(\ref{eq:e-h_propagator}) can also be given as
\begin{eqnarray*}
D_{e}\left(x,y\right) & = & \Theta\left(t_{x}-t_{y}\right)D_{e}^{>}\left(x,y\right)+\Theta\left(t_{y}-t_{x}\right)D_{e}^{<}\left(x,y\right),
\end{eqnarray*}
 where
\begin{eqnarray}
D_{e}^{>}\left(x,y\right) & = & \sum_{k>F}u_{k}\left(\mathbf{x}\right)u_{k}^{*}\left(\mathbf{y}\right)e^{-i\varepsilon_{k}\left(t_{x}-t_{y}\right)},
\end{eqnarray}
 and 
\begin{eqnarray}
D_{e}^{<}\left(x,y\right) & = & -\sum_{k\leq F}u_{k}\left(\mathbf{x}\right)u_{k}^{*}\left(\mathbf{y}\right)e^{-i\varepsilon_{k}\left(t_{x}-t_{y}\right)}.
\end{eqnarray}
Likewise, the propagator of DM particle is 
\begin{eqnarray}
D_{\chi}\left(x,y\right) & = & \int\sum\limits_{\mathbf{p}_{\chi}}\frac{i}{\omega-\varepsilon_{\mathbf{p}_{\chi}}+i0^{+}}\frac{e^{i\mathbf{p}_{\chi}\cdot\left(\mathbf{x}-\mathbf{y}\right)}}{V}\frac{e^{-i\omega\left(t_{x}-t_{y}\right)}\mathrm{d}\omega}{2\pi}\nonumber \\
 & = & \Theta\left(t_{x}-t_{y}\right)D_{\chi}^{>}\left(x,y\right)+\Theta\left(t_{y}-t_{x}\right)D_{\chi}^{<}\left(x,y\right),
\end{eqnarray}
where $\varepsilon_{\mathbf{p}_{\chi}}=\left|\mathbf{p}_{\chi}\right|^{2}/\left(2m_{\chi}\right)$,
\begin{eqnarray}
D_{\chi}^{>}\left(x,y\right) & = & \sum\limits_{\mathbf{p}_{\chi}}\frac{e^{i\mathbf{p}_{\chi}\cdot\left(\mathbf{x}-\mathbf{y}\right)}}{V}e^{-i\varepsilon_{\mathbf{p}}\left(t_{x}-t_{y}\right)},
\end{eqnarray}
and 
\begin{eqnarray}
D_{\chi}^{<}\left(x,y\right) & = & 0.
\end{eqnarray}
The difference between the DM particle propagator and the electron-hole
propagator lies in the fact that the ground state energy of the DM
particle is $0$. Since we are interested in the application of cutting
rules to the DM self-energy, here we focus on the Feynman diagram
$G\left(z,z'\right)$ in Eq.~(\ref{eq:G_function}), which is amputated
of its DM external legs. If its internal vertices $\left\{ y_{1},\cdots,y_{n},\cdots\right\} $
are explicitly considered, it can be further expressed as $G\left(z,z';y_{1},\cdots,y_{n},\cdots\right)$
in Eq.~(\ref{eq:G_function1}). For a more general discussion, we
adopt a unified notation $\left\{ x_{k}\right\} $ to represent the
original coordinates $\left\{ z,z';y_{1},\cdots,y_{n},\cdots\right\} $.
Then, one might expect that a straightforward application of the largest
time equation in the previous section can readily lead us to the cutting
rule in solids.

However, a simple observation suggests that the underlining rules
from the preceding section do not work for the DM and electron-hole
propagators, due to the fact that $D_{e}^{>}\left(x,y\right)^{*}\neq D_{e}^{<}\left(x,y\right)$.
Notice that the key support of the largest time equation is encoded
in Eqs. (\ref{eq:Ltime1}-\ref{eq:Ltime2}). These equations ensure
that underlining the coordinate with the largest time does not alter
the propagators connected to this coordinate. Consequently, the only
effect introduced by the underlining is an overall minus sign, arising
from the change $-i\lambda\rightarrow i\lambda$. But these properties
no longer hold in our present situation.

Therefore, in order to derive the cutting rules for DM and electrons
in solids, we need to find an alternative set of underlining rules.
To this end, we begin by introducing a new underlining operation on
the vertices in $G\left(x_{1},\cdots,x_{i},\cdots\right)$, which
are summarized as the following,
\begin{enumerate}
\item If the vertices $x_{i}$ and $x_{j}$ connected by the propagator
$D_{e\,\left(\chi\right)}\left(x_{i},x_{j}\right)$ are not underlined,
one keeps the original propagator $D_{e\,\left(\chi\right)}\left(x_{i},x_{j}\right)$.
\item If $x_{i}$ is underlined while $x_{j}$ is not underlined, the two
vertices are linked by $D_{e\,\left(\chi\right)}^{>}\left(x_{i},x_{j}\right).$ 
\item If $x_{j}$ is underlined while $x_{i}$ is not underlined, the two
vertices are linked by $D_{e\,\left(\chi\right)}^{<}\left(x_{i},x_{j}\right).$
\item If both $x_{i}$ and $x_{j}$ are underlined, one adopts the complex
conjugate of the reversed propagator $D_{e\,\left(\chi\right)}\left(x_{j},x_{i}\right)^{*}$.
\item If one of the vertices (say, $x_{i}$ for instance) of an instantaneous
Coulomb (Yukawa) propagator $D^{\mathrm{Cou}}\left(x_{i},x_{j}\right)$
($D^{A'}\left(x_{i},x_{j}\right)$) is underlined, the other must
be underlined, too, and $D^{\mathrm{Cou}}\left(x_{i},x_{j}\right)$
($D^{A'}\left(x_{i},x_{j}\right)$) is substituted by its complex
conjugate $D^{\mathrm{Cou}}\left(x_{i},x_{j}\right)^{*}$ ($D^{A'}\left(x_{i},x_{j}\right)^{*}$). 
\item If $x_{i}$ is underlined, the coupling strength factor $-ie$ (or
$ig_{\chi\left(e\right)}$) is changed to $ie$ (or $-ig_{\chi\left(e\right)}$). 
\end{enumerate}
Rules (2-6) can be summarized below, \begin{subequations}
\begin{align}
D_{e\,\left(\chi\right)}\left(\underline{x}_{i},x_{j}\right)&=D_{e\,\left(\chi\right)}^{>}\left(x_{i},x_{j}\right)\\
D_{e\,\left(\chi\right)}\left(x_{i},\underline{x}_{j}\right)&=D_{e\,\left(\chi\right)}^{<}\left(x_{i},x_{j}\right)\\
D_{e\,\left(\chi\right)}\left(\underline{x}_{i},\underline{x}_{j}\right)&=D_{e\,\left(\chi\right)}\left(x_{j},x_{i}\right)^{*}\label{eq:fermion*}\\
D^{\mathrm{Cou}\,\left(A'\right)}\left(\underline{x}_{i},\underline{x}_{j}\right)&=D^{\mathrm{Cou}\,\left(A'\right)}\left(x_{i},x_{j}\right)^{*}\label{eq:Coulomb1*}.
\end{align}
\end{subequations}Using a circle to represent the underlined vertex, these rules can
be shown diagrammatically as follows,\begin{subequations}
\begin{align}
\vcenter{\hbox{\tikzset{every picture/.style={line width=0.75pt}}
\begin{tikzpicture}[x=0.55pt,y=0.55pt,yscale=-1,baseline={([yshift=20pt]current bounding box.center)}]
\begin{scope}[shift={(150,100)}]
\draw[line width=1.2pt,
    decoration={markings, mark=at position 0.57 with {\arrow{Triangle[scale=0.65]}}},
    postaction={decorate}
  ] 
  (100,0) -- (0,0) ;
  \filldraw[black] (0,0) circle[radius=1.2pt]
    node[below=2pt] {$x_{i}$};
  \filldraw[black] (100,0) circle[radius=1.2pt]
    node[below=2pt] {$x_{j}$};
\draw[line width=1.2pt] (0,0) circle[radius=6.5];
\end{scope}
\end{tikzpicture}}} 
&= D_{e}^{>}\left(x_{i},x_{j}\right)=\sum_{k>F}u_{k}\left(\mathbf{x}_{i}\right)u_{k}^{*}\left(\mathbf{x}_{j}\right)e^{-i\varepsilon_{k}\left(t_{i}-t_{j}\right)},
\label{eq:fermion3*}\\
\vcenter{\hbox{\tikzset{every picture/.style={line width=0.75pt}}
\begin{tikzpicture}[x=0.55pt,y=0.55pt,yscale=-1,baseline={([yshift=20pt]current bounding box.center)}]
\begin{scope}[shift={(150,100)}]
\draw[line width=1.2pt,
    decoration={markings, mark=at position 0.57 with {\arrow{Triangle[scale=0.65]}}},
    postaction={decorate}
  ] 
  (100,0) -- (0,0) ;
  \filldraw[black] (0,0) circle[radius=1.2pt]
    node[below=2pt] {$x_{i}$};
  \filldraw[black] (100,0) circle[radius=1.2pt]
    node[below=2pt] {$x_{j}$};
\draw[line width=1.2pt] (100,0) circle[radius=6.5];
\end{scope}
\end{tikzpicture}}} 
&= D_{e}^{<}\left(x_{i},x_{j}\right)=-\sum_{k\leq F}u_{k}\left(\mathbf{x}_{i}\right)u_{k}^{*}\left(\mathbf{x}_{j}\right)e^{-i\varepsilon_{k}\left(t_{i}-t_{j}\right)},\\
\vcenter{\hbox{\tikzset{every picture/.style={line width=0.75pt}}
\begin{tikzpicture}[x=0.55pt,y=0.55pt,yscale=-1,baseline={([yshift=20pt]current bounding box.center)}]
\begin{scope}[shift={(150,100)}]
\draw[line width=1.2pt,
    decoration={markings, mark=at position 0.57 with {\arrow{Triangle[scale=0.65]}}},
    postaction={decorate}
  ] 
  (100,0) -- (0,0) ;
  \filldraw[black] (0,0) circle[radius=1.2pt]
    node[below=2pt] {$x_{i}$};
  \filldraw[black] (100,0) circle[radius=1.2pt]
    node[below=2pt] {$x_{j}$};
\draw[line width=1.2pt] (0,0) circle[radius=6.5];
\draw[line width=1.2pt] (100,0) circle[radius=6.5];
\end{scope}
\end{tikzpicture}}} 
&=\left(
\vcenter{\hbox{\tikzset{every picture/.style={line width=0.75pt}}
\begin{tikzpicture}[x=0.55pt,y=0.55pt,yscale=-1,baseline={([yshift=20pt]current bounding box.center)}]
\begin{scope}[shift={(150,100)}]
\draw[line width=1.2pt,
    decoration={markings, mark=at position 0.57 with {\arrow{Triangle[scale=0.65]}}},
    postaction={decorate}
  ] 
  (0,0) -- (100,0) ;
  \filldraw[black] (0,0) circle[radius=1.2pt]
    node[below=2pt] {$x_{i}$};
  \filldraw[black] (100,0) circle[radius=1.2pt]
    node[below=2pt] {$x_{j}$};
\end{scope}
\end{tikzpicture}}}
\right)^{*}
=D_{e}\left(x_{j},x_{i}\right)^{*}\label{eq:fermion2*},
\\
\vcenter{\hbox{\tikzset{every picture/.style={line width=0.75pt}}
\begin{tikzpicture}[x=0.55pt,y=0.55pt,yscale=-1,baseline={([yshift=20pt]current bounding box.center)}]
\begin{scope}[shift={(150,100)}]
\draw[gray!70,line width=1.2pt,
    decoration={markings, mark=at position 0.57 with {\arrow{Triangle[scale=0.65]}}},
    postaction={decorate}
  ] 
  (100,0) -- (0,0) ;
  \filldraw[black] (0,0) circle[radius=1.2pt]
    node[below=2pt] {$x_{i}$};
  \filldraw[black] (100,0) circle[radius=1.2pt]
    node[below=2pt] {$x_{j}$};
\draw[line width=1.2pt] (0,0) circle[radius=6.5];
\end{scope}
\end{tikzpicture}}} 
&= D_{\chi}^{>}\left(x_{i},x_{j}\right)=\sum\limits_{\mathbf{p_{\chi}}}\frac{e^{i\mathbf{p_{\chi}}\cdot\left(\mathbf{x}_{i}-\mathbf{x}_{j}\right)}}{V}e^{-i\varepsilon_{\mathbf{p_{\chi}}}\left(t_{i}-t_{j}\right)},\\
\vcenter{\hbox{\tikzset{every picture/.style={line width=0.75pt}}
\begin{tikzpicture}[x=0.55pt,y=0.55pt,yscale=-1,baseline={([yshift=20pt]current bounding box.center)}]
\begin{scope}[shift={(150,100)}]
\draw[gray!70,line width=1.2pt,
    decoration={markings, mark=at position 0.57 with {\arrow{Triangle[scale=0.65]}}},
    postaction={decorate}
  ] 
  (100,0) -- (0,0) ;
  \filldraw[black] (0,0) circle[radius=1.2pt]
    node[below=2pt] {$x_{i}$};
  \filldraw[black] (100,0) circle[radius=1.2pt]
    node[below=2pt] {$x_{j}$};
\draw[line width=1.2pt] (100,0) circle[radius=6.5];
\end{scope}
\end{tikzpicture}}} 
&= D_{\chi}^{<}\left(x_{i},x_{j}\right)=0\label{eq:DM_0},\\
\vcenter{\hbox{\tikzset{every picture/.style={line width=0.75pt}}
\begin{tikzpicture}[x=0.55pt,y=0.55pt,yscale=-1,baseline={([yshift=20pt]current bounding box.center)}]
\begin{scope}[shift={(150,100)}]
\draw[gray!70,line width=1.2pt,
    decoration={markings, mark=at position 0.57 with {\arrow{Triangle[scale=0.65]}}},
    postaction={decorate}
  ] 
  (100,0) -- (0,0) ;
  \filldraw[black] (0,0) circle[radius=1.2pt]
    node[below=2pt] {$x_{i}$};
  \filldraw[black] (100,0) circle[radius=1.2pt]
    node[below=2pt] {$x_{j}$};
\draw[line width=1.2pt] (0,0) circle[radius=6.5];
\draw[line width=1.2pt] (100,0) circle[radius=6.5];
\end{scope}
\end{tikzpicture}}} 
&=\left(
\vcenter{\hbox{\tikzset{every picture/.style={line width=0.75pt}}
\begin{tikzpicture}[x=0.55pt,y=0.55pt,yscale=-1,baseline={([yshift=20pt]current bounding box.center)}]
\begin{scope}[shift={(150,100)}]
\draw[gray!70,line width=1.2pt,
    decoration={markings, mark=at position 0.57 with {\arrow{Triangle[scale=0.65]}}},
    postaction={decorate}
  ] 
  (0,0) -- (100,0) ;
  \filldraw[black] (0,0) circle[radius=1.2pt]
    node[below=2pt] {$x_{i}$};
  \filldraw[black] (100,0) circle[radius=1.2pt]
    node[below=2pt] {$x_{j}$};
\end{scope}
\end{tikzpicture}}}
\right)^{*}
=D_{\chi}\left(x_{j},x_{i}\right)^{*}\label{eq:DMfermion2*},
\\
\vcenter{\hbox{\tikzset{every picture/.style={line width=0.75pt}}
\begin{tikzpicture}[x=0.55pt,y=0.55pt,yscale=-1,baseline={([yshift=20pt]current bounding box.center)}]
\begin{scope}[shift={(150,100)}]
\draw[line width=1.2pt,
    decorate,
    decoration={snake, 
               amplitude=2pt,    
               segment length=8pt}] 
  (100,0) -- (0,0);
  \filldraw[black] (0,0) circle[radius=1.2pt]
    node[below=2pt] {$x_{i}$};
  \filldraw[black] (100,0) circle[radius=1.2pt]
    node[below=2pt] {$x_{j}$};
\draw[line width=1.2pt] (0,0) circle[radius=6.5];
\draw[line width=1.2pt] (100,0) circle[radius=6.5];
\end{scope}
\end{tikzpicture}}} 
&= \left(
\vcenter{\hbox{\tikzset{every picture/.style={line width=0.75pt}}
\begin{tikzpicture}[x=0.55pt,y=0.55pt,yscale=-1,baseline={([yshift=20pt]current bounding box.center)}]
\begin{scope}[shift={(150,100)}]
\draw[line width=1.2pt,
    decorate,
    decoration={snake, 
               amplitude=2pt,    
               segment length=8pt}] 
  (100,0) -- (0,0);
  \filldraw[black] (0,0) circle[radius=1.2pt]
    node[below=2pt] {$x_{i}$};
  \filldraw[black] (100,0) circle[radius=1.2pt]
    node[below=2pt] {$x_{j}$};
\end{scope}
\end{tikzpicture}}}
\right)^{*}=D^{\mathrm{Cou}}\left(x_{i},x_{j}\right)^{*}=-D^{\mathrm{Cou}}\left(x_{i},x_{j}\right)\label{eq:Coulomb2*},\\
\vcenter{\hbox{\tikzset{every picture/.style={line width=0.75pt}}
\begin{tikzpicture}[x=0.55pt,y=0.55pt,yscale=-1,baseline={([yshift=20pt]current bounding box.center)}]
\begin{scope}[shift={(150,100)}]
\draw[line width=1.2pt,dashed] (100,0) -- (0,0);
\filldraw[black] (0,0) circle[radius=1.2pt]
    node[below=2pt] {$x_{i}$};
  \filldraw[black] (100,0) circle[radius=1.2pt]
    node[below=2pt] {$x_{j}$};
\draw[line width=1.2pt] (0,0) circle[radius=6.5];
\draw[line width=1.2pt] (100,0) circle[radius=6.5];
\end{scope}
\end{tikzpicture}}} 
&= \left(
\vcenter{\hbox{\tikzset{every picture/.style={line width=0.75pt}}
\begin{tikzpicture}[x=0.55pt,y=0.55pt,yscale=-1,baseline={([yshift=20pt]current bounding box.center)}]
\begin{scope}[shift={(150,100)}]
\draw[line width=1.2pt,dashed] (100,0) -- (0,0);
\filldraw[black] (0,0) circle[radius=1.2pt]
    node[below=2pt] {$x_{i}$};
  \filldraw[black] (100,0) circle[radius=1.2pt]
    node[below=2pt] {$x_{j}$};
\end{scope}
\end{tikzpicture}}} 
\right)^{*}=D^{A'}\left(x_{i},x_{j}\right)^{*}=-D^{A'}\left(x_{i},x_{j}\right)\label{eq:dark_photon*}.
\end{align}
\end{subequations}

Then we are able to also establish a largest time equation analogous
to the one in relativistic scalar field. Suppose in $G\left(x_{1},\cdots,x_{i},\cdots,x_{n+2}\right)$
the time coordinate $x_{i}^{0}$ is larger than any other $x_{j}^{0}$,
one has \begin{subequations}
\begin{align}
D_{e\,\left(\chi\right)}\left(\underline{x}_{i},x_{j}\right)&=D_{e\,\left(\chi\right)}^{>}\left(x_{i},x_{j}\right)=D_{e\,\left(\chi\right)}\left(x_{i},x_{j}\right),\\
D_{e\,\left(\chi\right)}\left(x_{j},\underline{x}_{i}\right)&=D_{e\,\left(\chi\right)}^{<}\left(x_{j},x_{i}\right)=D_{e\,\left(\chi\right)}\left(x_{j},x_{i}\right),\\
D_{e\,\left(\chi\right)}\left(\underline{x}_{i},\underline{x}_{j}\right)&=D_{e\,\left(\chi\right)}\left(x_{j},x_{i}\right)^{*}=D_{e\,\left(\chi\right)}^{<}\left(x_{j},x_{i}\right)^{*}=D_{e\,\left(\chi\right)}^{<}\left(x_{i},x_{j}\right)=D_{e\,\left(\chi\right)}\left(x_{i},\underline{x}_{j}\right),\\
D_{e\,\left(\chi\right)}\left(\underline{x}_{j},\underline{x}_{i}\right)&=D_{e\,\left(\chi\right)}\left(x_{i},x_{j}\right)^{*}=D_{e\,\left(\chi\right)}^{>}\left(x_{i},x_{j}\right)^{*}=D_{e\,\left(\chi\right)}^{>}\left(x_{j},x_{i}\right)=D_{e\,\left(\chi\right)}\left(\underline{x}_{j},x_{i}\right),\\
D^{\mathrm{Cou}\,\left(A'\right)}\left(\underline{x}_{i},\underline{x}_{j}\right)&=D^{\mathrm{Cou}\,\left(A'\right)}\left(x_{i},x_{j}\right)^{*}=-D^{\mathrm{Cou}\,\left(A'\right)}\left(x_{i},x_{j}\right).
\end{align}
\end{subequations}It is observed that underlining $x_{i}$ with the largest time component
leaves the electron propagator unchanged but introduces a minus sign
for the Coulomb and the Yukawa propagators. Additionally, the two
minus signs arising from the replacements $-ie\rightarrow ie$ ($ig_{\chi\left(e\right)}\rightarrow-ig_{\chi\left(e\right)}$)
for the pair-underlined vertices $\underline{x}_{i}$ and $\underline{x}_{j}$
of the Coulomb (massive mediator) internal line neutralize each other,
and hence one has $G\left(x_{1},\cdots,\underline{x}_{i},\cdots,\underline{x}_{j},\cdots,x_{n},\cdots\right)=-G\left(x_{1},\cdots,x_{i},\cdots,x_{j},\cdots,x_{n},\cdots\right)$.
For a general $G$ with arbitrary coordinate configuration, say, $G\left(\underline{x}_{1},\cdots,x_{i},\cdots,x_{j},\cdots,x_{n},\cdots\right)$,
where some coordinates are underlined, we assume the coordinate $x_{i}$
has the largest time component and there also exists a vertex $x_{j}$
that connects $x_{i}$ via instantaneous Coulomb interaction or DM-electron
interaction. Therefore, if $x_{i}$ is underlined, $x_{j}$ must also
be underlined, leading to
\begin{eqnarray}
G\left(\underline{x}_{1},\cdots,\underline{x}_{i},\cdots,\underline{x}_{j},\cdots,x_{n},\cdots\right)+G\left(\underline{x}_{1},\cdots,x_{i},\cdots,x_{j},\cdots,x_{n},\cdots\right) & = & 0.
\end{eqnarray}
This relation holds for any other coordinate configurations, yielding
\begin{eqnarray}
\sum G\left(x_{1},\cdots,\underline{x}_{k},\cdots,x_{n},\cdots\right) & = & 0,\label{eq:sum_zero}
\end{eqnarray}
 where the sum is over all possibilities of underlining all vertices.
This constitutes the first part of our proof of the cutting rules.

We now try to establish a connection between the imaginary part of
the amplitude in Eq.~(\ref{eq:decay_rate}) and the quantity $G\left(\underline{x}_{1},\cdots,\underline{x}_{i},\cdots,\underline{x}_{n},\cdots\right)$,
in which all coordinates are underlined. To achieve this, we start
from a simple example,\emph{ i.e.}, we use a fermion loop \begin{tikzpicture}[baseline={(0,0.4)},scale=0.6,
  thick,
  fermion/.style={
    postaction={decorate},
    decoration={
      markings,
      mark=at position 0.57 with {\arrow{Triangle[scale=0.65, reversed]}}
    },
    thick
  },
fermionloop/.style={
    postaction={decorate},
    decoration={
      markings},
    thick},
momentum/.style={
    postaction={decorate},
    decoration={
      markings,
      mark=at position 0.28 with {\arrow{Triangle[scale=0.65]}},
      mark=at position 0.8 with {\arrow{Triangle[scale=0.65]}},
    }
  }]

\coordinate (A) at (-2,0);
\coordinate (B) at (-1,0);
\coordinate (C) at (1,0);
\coordinate (D) at (2,0);
\coordinate (X) at (0,0.9); 

\filldraw[lightgray!50] (X) circle[radius=0.5cm]; 
\draw[fermionloop,line width=1.2pt] (X)  circle[radius=0.5cm];
\path[momentum] (X)  circle[radius=0.5cm];
\filldraw[fill=black] (-0.5,0.84)node[left=1.5pt] {\small$y$} circle[radius=1.2pt];
\filldraw[fill=black] (0.5,0.84) node[right=1.5pt] {\small$y'$} circle[radius=1.2pt];
\end{tikzpicture},
 to exemplify the diagram \blobFigure[baseline={(0,0.4)},scale=0.6]
in Eq.~(\ref{eq:G_function1}). This loop may contain more intricate
substructures, with its two vertices $y$ and $y'$ coupled to massive
mediator propagators. When all vertices are underlined, the electron-hole
propagator has the property $D\left(\underline{x}_{i},\underline{x}_{j}\right)=D\left(x_{j},x_{i}\right)^{*}$
(see Eq.~(\ref{eq:fermion*})), which means that when both ends are
underlined, the electron propagator not only becomes its complex conjugate,
but the momentum flow is also reversed ( Eq.~(\ref{eq:fermion2*})),
while each Coulomb line (see Eq.~(\ref{eq:Coulomb2*})) and the coupling
at every vertex becomes their complex conjugates ($-ie\rightarrow ie$
($ig_{\chi\left(e\right)}\rightarrow-ig_{\chi\left(e\right)}$)).
Or equivalently, one can turn over the coordinates along the loop
while keeping the momentum flow unchanged, which leads to 

\begin{equation}
\vcenter{\hbox{\tikzset{every picture/.style={line width=0.75pt}}
\begin{tikzpicture}[
  thick,
  fermion/.style={draw=black,
    postaction={decorate},
    decoration={
      markings,
      mark=at position 0.57 with {\arrow{Triangle[scale=0.65,reversed]}}
    },
    thick
  },
 photon/.style={
  thick,
  decorate,                  
  decoration={
    snake,                   
    amplitude=1.5pt,           
    segment length=5pt       
  }
}]

\coordinate (y1) at (-1.8,0);
\coordinate (x1) at (-1.2,0);
\coordinate (x2) at (-0.6,0);
\coordinate (xn_1) at (0.6,0);
\coordinate (xn) at (1.2,0);
\coordinate (y2) at (1.8,0);
\coordinate (y1h) at (-1.8,0.5);
\coordinate (x1h) at (-1.2,0.5);
\coordinate (x2h) at (-0.6,0.5);
\coordinate (xn_1h) at (0.6,0.5);
\coordinate (xnh) at (1.2,0.5);
\coordinate (y2h) at (1.8,0.5);

\draw[fermion,line width=1.2pt] (y1)node[left] {$\cdots$} node[below=2pt] {$y$} -- (x1) node[below=2pt] {$x_{1}$};
\draw[fermion,line width=1.2pt] (x1) -- (x2) node[below=2pt] {$x_{2}$} node[right] {$\cdots\,\cdots$};
\draw[fermion,line width=1.2pt] (xn_1)node[below=2pt] {$x_{n-1}$} -- (xn)  node[below=2pt] {$x_{n}$};
\draw[fermion,line width=1.2pt] (xn) -- (y2) node[below=2pt] {$y'$} node[right] {$\cdots$};
\filldraw[fill=black] (y1) circle[radius=1.2pt];
\filldraw[fill=black] (x1) circle[radius=1.2pt];
\filldraw[fill=black] (x2) circle[radius=1.2pt];
\filldraw[fill=black] (xn_1) circle[radius=1.2pt];
\filldraw[fill=black] (xn) circle[radius=1.2pt];
\filldraw[fill=black] (y2) circle[radius=1.2pt];
\draw[photon,line width=1.2pt] (y1) --(y1h);
\draw[photon,line width=1.2pt] (x1) --(x1h);
\draw[photon,line width=1.2pt] (x2) --(x2h);
\draw[photon,line width=1.2pt] (xn_1) --(xn_1h);
\draw[photon,line width=1.2pt] (xn) --(xnh);
\draw[photon,line width=1.2pt] (y2) --(y2h);
\draw[line width=1.2pt] (y1) circle[radius=0.12];
\draw[line width=1.2pt] (x1) circle[radius=0.12];
\draw[line width=1.2pt] (x2) circle[radius=0.12];
\draw[line width=1.2pt] (xn_1) circle[radius=0.12];
\draw[line width=1.2pt] (xn) circle[radius=0.12];
\draw[line width=1.2pt] (y2) circle[radius=0.12];
\end{tikzpicture}}}=
\left(
\vcenter{\hbox{\tikzset{every picture/.style={line width=0.75pt}}
\begin{tikzpicture}[
  thick,
  fermion/.style={draw=black,
    postaction={decorate},
    decoration={
      markings,
      mark=at position 0.57 with {\arrow{Triangle[scale=0.65,reversed]}}
    },
    thick
  },
 photon/.style={
  thick,
  decorate,                  
  decoration={
    snake,                   
    amplitude=1.5pt,           
    segment length=5pt}       
  }]

\coordinate (y1) at (-1.8,0);
\coordinate (x1) at (-1.2,0);
\coordinate (x2) at (-0.6,0);
\coordinate (xn_1) at (0.6,0);
\coordinate (xn) at (1.2,0);
\coordinate (y2) at (1.8,0);
\coordinate (y1h) at (-1.8,0.5);
\coordinate (x1h) at (-1.2,0.5);
\coordinate (x2h) at (-0.6,0.5);
\coordinate (xn_1h) at (0.6,0.5);
\coordinate (xnh) at (1.2,0.5);
\coordinate (y2h) at (1.8,0.5);

\draw[fermion,line width=1.2pt] (y1)node[left] {$\cdots$} node[below=2pt] {$y'$} -- (x1) node[below=2pt] {$x_{n}$};
\draw[fermion,line width=1.2pt] (x1) -- (x2) node[below=2pt] {$x_{n-1}$} node[right] {$\cdots\,\cdots$};
\draw[fermion,line width=1.2pt] (xn_1)node[below=2pt] {$x_{2}$} -- (xn)  node[below=2pt] {$x_{1}$};
\draw[fermion,line width=1.2pt] (xn) -- (y2) node[below=2pt] {$y$} node[right] {$\cdots$};
\filldraw[fill=black] (y1) circle[radius=1.2pt];
\filldraw[fill=black] (x1) circle[radius=1.2pt];
\filldraw[fill=black] (x2) circle[radius=1.2pt];
\filldraw[fill=black] (xn_1) circle[radius=1.2pt];
\filldraw[fill=black] (xn) circle[radius=1.2pt];
\filldraw[fill=black] (y2) circle[radius=1.2pt];
\draw[photon,line width=1.2pt] (y1) --(y1h);
\draw[photon,line width=1.2pt] (x1) --(x1h);
\draw[photon,line width=1.2pt] (x2) --(x2h);
\draw[photon,line width=1.2pt] (xn_1) --(xn_1h);
\draw[photon,line width=1.2pt] (xn) --(xnh);
\draw[photon,line width=1.2pt] (y2) --(y2h);

\end{tikzpicture}}}
\right)^{*},
\end{equation}with $\left\{ x_{k}\right\} $ representing internal vertices located
on the outer fermion loop. Then the following relation holds: \begin{equation}
\vcenter{\hbox{\tikzset{every picture/.style={line width=0.75pt}}
\begin{tikzpicture}[
  thick,
  fermion/.style={
    postaction={decorate},
    decoration={
      markings,
      mark=at position 0.57 with {\arrow{Triangle[scale=0.65, reversed]}}
    },
    thick
  },
fermionloop/.style={
    postaction={decorate},
    decoration={
      markings},
    thick},
momentum/.style={
    postaction={decorate},
    decoration={
      markings,
      mark=at position 0.28 with {\arrow{Triangle[scale=0.65]}},
      mark=at position 0.78 with {\arrow{Triangle[scale=0.65]}},
    }
  }]

\coordinate (A) at (-2,0);
\coordinate (B) at (-1,0);
\coordinate (C) at (1,0);
\coordinate (D) at (2,0);
\coordinate (X) at (0,0.9); 

\filldraw[lightgray!50] (X) circle[radius=0.5cm]; 
\draw[fermionloop,line width=1.2pt] (X)  circle[radius=0.5cm];
\path[momentum] (X)  circle[radius=0.5cm];
\filldraw[fill=black] (-0.5,0.84)node[left=1.5pt] {\small$y$} circle[radius=1.2pt];
\filldraw[fill=black] (0.5,0.84) node[right=1.5pt] {\small$y'$} circle[radius=1.2pt];
\draw[line width=1.2pt] (-0.5,0.84) circle[radius=0.12];
\draw[line width=1.2pt] (0.5,0.84) circle[radius=0.12];
\end{tikzpicture}}}=
\left(
\vcenter{\hbox{\tikzset{every picture/.style={line width=0.75pt}}
\begin{tikzpicture}[
  thick,
  fermion/.style={
    postaction={decorate},
    decoration={
      markings,
      mark=at position 0.57 with {\arrow{Triangle[scale=0.65, reversed]}}
    },
    thick
  },
fermionloop/.style={
    postaction={decorate},
    decoration={
      markings},
    thick},
momentum/.style={
    postaction={decorate},
    decoration={
      markings,
      mark=at position 0.28 with {\arrow{Triangle[scale=0.65]}},
      mark=at position 0.78 with {\arrow{Triangle[scale=0.65]}},
    }
  }]

\coordinate (A) at (-2,0);
\coordinate (B) at (-1,0);
\coordinate (C) at (1,0);
\coordinate (D) at (2,0);
\coordinate (X) at (0,0.9); 

\filldraw[lightgray!50] (X) circle[radius=0.5cm]; 
\draw[fermionloop,line width=1.2pt] (X)  circle[radius=0.5cm];
\path[momentum] (X)  circle[radius=0.5cm];
\filldraw[fill=black] (-0.5,0.84)node[left=1.5pt] {\small$y'$} circle[radius=1.2pt];
\filldraw[fill=black] (0.5,0.84) node[right=1.5pt] {\small$y$} circle[radius=1.2pt];
\draw[dashed] (0,1.25) -- (0,0.45); 
\draw[<-] (0.25,0.99) -- (-0.25,0.99);
\draw[->] (0.25,0.69) -- (-0.25,0.69); 
\end{tikzpicture}}}
\right)^{*},
\end{equation}
where all vertices on the LHS are underlined, while the positions
of internal vertices on the RHS are flipped left-to-right. Combining
this relation and the underlining rules in Eq.~(\ref{eq:DMfermion2*})
for DM particle, as well as Eq.~(\ref{eq:dark_photon*}) for massive
mediator, we obtain a generic expression for $G\left(\underline{z},\underline{z}';\underline{y},\underline{y}',\cdots\right)$
in the original coordinate notation, that is, \begin{equation}
G\left(\underline{z},\underline{z}';\underline{y},\underline{y}',\cdots\right)=\vcenter{\hbox{\tikzset{every picture/.style={line width=0.75pt}}
\begin{tikzpicture}[
  thick,
  DMfermion/.style={draw=gray!70,
    postaction={decorate},
    decoration={
      markings,
      mark=at position 0.52 with {\arrow{Triangle[scale=0.65, reversed]}}
    },
    thick
  },
  fermion/.style={
    postaction={decorate},
    decoration={
      markings,
      mark=at position 0.57 with {\arrow{Triangle[scale=0.65, reversed]}}
    },
    thick
  },
  photon/.style={
    dashed,          
    thick
  },
fermionloop/.style={
    postaction={decorate},
    decoration={
      markings},
    thick},
   sparse crosshatch/.style={
    pattern={Hatch[angle=45,distance=4.2pt,line width=0.7pt]},
    pattern color=black
  },
momentum/.style={
    postaction={decorate},
    decoration={
      markings,
      mark=at position 0.28 with {\arrow{Triangle[scale=0.65]}},
      mark=at position 0.78 with {\arrow{Triangle[scale=0.65]}},
    }
  }]

\coordinate (A) at (-2,0);
\coordinate (B) at (-1,0);
\coordinate (C) at (1,0);
\coordinate (D) at (1,0);
\coordinate (X) at (0,0.8); 
\coordinate (aleft) at (-0.5,1.4); 
\coordinate (aright) at (0.5,1.4);
\coordinate (yleft) at (-1,0.85);
\coordinate (yright) at (1,0.85);
\draw[DMfermion,line width=1.2pt] (B) -- (C) node[midway,below] {$$} node[right] {};
\filldraw[fill=black] (B) circle[radius=1.2pt];
\filldraw[fill=black] (C) circle[radius=1.2pt];

\draw[photon,line width=1.2pt] (B) to[out=90,in=180] coordinate[pos=0.7] (X1) (X);
\draw[photon,line width=1.2pt] (X) to[out=0,in=90] coordinate[pos=0.7] (X2) (C);
\filldraw[lightgray!50] (X) circle[radius=0.5cm]; 
\draw[fermionloop,line width=1.2pt] (X)  circle[radius=0.5cm];
\path[momentum] (X)  circle[radius=0.5cm];
\filldraw[fill=black] (-0.49,0.67)node[left=5.5pt] {\small$y$} circle[radius=1.2pt];
\filldraw[fill=black] (0.49,0.67) node[right=5.5pt] {\small$y'$} circle[radius=1.2pt];
\draw[line width=1.2pt] (B) circle[radius=0.12];
\draw[line width=1.2pt] (C) circle[radius=0.12];
\draw[line width=1.2pt] (-0.49,0.67) circle[radius=0.12];
\draw[line width=1.2pt] (0.49,0.67) circle[radius=0.12];
\node at(B)[left=4pt] {$z$};
\node at(C)[right=4pt] {$z'$};
\end{tikzpicture}}}=\\
\left(\vcenter{\hbox{\tikzset{every picture/.style={line width=0.75pt}}
\begin{tikzpicture}[
  thick,
  DMfermion/.style={draw=gray!70,
    postaction={decorate},
    decoration={
      markings,
      mark=at position 0.52 with {\arrow{Triangle[scale=0.65, reversed]}}
    },
    thick
  },
  fermion/.style={
    postaction={decorate},
    decoration={
      markings,
      mark=at position 0.57 with {\arrow{Triangle[scale=0.65, reversed]}}
    },
    thick
  },
  photon/.style={
    dashed,          
    thick
  },
fermionloop/.style={
    postaction={decorate},
    decoration={
      markings},
    thick},
   sparse crosshatch/.style={
    pattern={Hatch[angle=45,distance=4.2pt,line width=0.7pt]},
    pattern color=black
  },
momentum/.style={
    postaction={decorate},
    decoration={
      markings,
      mark=at position 0.28 with {\arrow{Triangle[scale=0.65]}},
      mark=at position 0.78 with {\arrow{Triangle[scale=0.65]}},
    }
  }]

\coordinate (A) at (-2,0);
\coordinate (B) at (-1,0);
\coordinate (C) at (1,0);
\coordinate (D) at (1,0);
\coordinate (X) at (0,0.8); 
\coordinate (aleft) at (-0.5,1.4); 
\coordinate (aright) at (0.5,1.4);
\coordinate (yleft) at (-1,0.85);
\coordinate (yright) at (1,0.85);
\draw[DMfermion,line width=1.2pt] (B) -- (C) node[midway,below] {$$} node[right] {};
\filldraw[fill=black] (B) circle[radius=1.2pt];
\filldraw[fill=black] (C) circle[radius=1.2pt];

\draw[photon,line width=1.2pt] (B) to[out=90,in=180] coordinate[pos=0.7] (X1) (X);
\draw[photon,line width=1.2pt] (X) to[out=0,in=90] coordinate[pos=0.7] (X2) (C);
\filldraw[lightgray!50] (X) circle[radius=0.5cm]; 
\draw[fermionloop,line width=1.2pt] (X)  circle[radius=0.5cm];
\path[momentum] (X)  circle[radius=0.5cm];
\filldraw[fill=black] (-0.49,0.67)node[left=5.5pt] {\small$y'$} circle[radius=1.2pt];
\filldraw[fill=black] (0.49,0.67) node[right=5.5pt] {\small$y$} circle[radius=1.2pt];
\node at(B)[left=4pt] {$z'$};
\node at(C)[right=4pt] {$z$};
\draw[dashed] (0,1.2) -- (0,0.4); 
\draw[<-] (0.25,0.95) -- (-0.25,0.95);
\draw[->] (0.25,0.65) -- (-0.25,0.65); 
\end{tikzpicture}}}
\right)^{*}=G\left(z',z;y',y,\cdots\right)^{*}.
\end{equation}Inserting this back into Eq.~(\ref{eq:sum_zero}) yields 

\begin{eqnarray}
G\left(z,z';y,y',\cdots\right)+G\left(z',z;y',y,\cdots\right)^{*} & = & -\sum_{\mathrm{partial}}G\left(z,\underline{z}';y,\underline{y}',\cdots\right),\label{eq:G_G_G}
\end{eqnarray}
where the sum on the RHS runs over all possible underlining vertices,
except for the two cases with no vertex or all vertices are underlined.
Since we cannot in general assert that $G\left(z',z;y',y,\cdots\right)=G\left(z,z';y,y',\cdots\right)$,
no equation analogous to Eq.~(\ref{eq:sum_zero-2}) is available
at this stage of the analysis. But if the gray fermion loop exhibits
the topological invariance under a flip operation,\emph{ i.e.}, \begin{equation}
\vcenter{\hbox{\tikzset{every picture/.style={line width=0.75pt}}
\begin{tikzpicture}[
  thick,
  fermion/.style={
    postaction={decorate},
    decoration={
      markings,
      mark=at position 0.57 with {\arrow{Triangle[scale=0.65, reversed]}}
    },
    thick
  },
fermionloop/.style={
    postaction={decorate},
    decoration={
      markings},
    thick},
momentum/.style={
    postaction={decorate},
    decoration={
      markings,
      mark=at position 0.28 with {\arrow{Triangle[scale=0.65]}},
      mark=at position 0.78 with {\arrow{Triangle[scale=0.65]}},
    }
  }]

\coordinate (A) at (-2,0);
\coordinate (B) at (-1,0);
\coordinate (C) at (1,0);
\coordinate (D) at (2,0);
\coordinate (X) at (0,0.9); 

\filldraw[lightgray!50] (X) circle[radius=0.5cm]; 
\draw[fermionloop,line width=1.2pt] (X)  circle[radius=0.5cm];
\path[momentum] (X)  circle[radius=0.5cm];
\filldraw[fill=black] (-0.5,0.84)node[left=1.5pt] {\small$y'$} circle[radius=1.2pt];
\filldraw[fill=black] (0.5,0.84) node[right=1.5pt] {\small$y$} circle[radius=1.2pt];
\draw[dashed] (0,1.25) -- (0,0.45); 
\draw[<-] (0.25,0.99) -- (-0.25,0.99);
\draw[->] (0.25,0.69) -- (-0.25,0.69); 
\end{tikzpicture}}}=
\vcenter{\hbox{\tikzset{every picture/.style={line width=0.75pt}}
\begin{tikzpicture}[
  thick,
  fermion/.style={
    postaction={decorate},
    decoration={
      markings,
      mark=at position 0.57 with {\arrow{Triangle[scale=0.65, reversed]}}
    },
    thick
  },
fermionloop/.style={
    postaction={decorate},
    decoration={
      markings},
    thick},
momentum/.style={
    postaction={decorate},
    decoration={
      markings,
      mark=at position 0.28 with {\arrow{Triangle[scale=0.65]}},
      mark=at position 0.78 with {\arrow{Triangle[scale=0.65]}},
    }
  }]

\coordinate (A) at (-2,0);
\coordinate (B) at (-1,0);
\coordinate (C) at (1,0);
\coordinate (D) at (2,0);
\coordinate (X) at (0,0.9); 

\filldraw[lightgray!50] (X) circle[radius=0.5cm]; 
\draw[fermionloop,line width=1.2pt] (X)  circle[radius=0.5cm];
\path[momentum] (X)  circle[radius=0.5cm];
\filldraw[fill=black] (-0.5,0.84)node[left=1.5pt] {\small$y'$} circle[radius=1.2pt];
\filldraw[fill=black] (0.5,0.84) node[right=1.5pt] {\small$y$} circle[radius=1.2pt];
\end{tikzpicture}}},
\end{equation} then convolving Eq.~(\ref{eq:G_G_G}) with outgoing and incoming
DM external legs shown in Eq.~(\ref{eq:G_function}) will give 
\begin{align}
& \left[i\mathcal{M}\left(p_{\chi}\rightarrow p_{\chi}\right)-i\mathcal{M}^{*}\left(p_{\chi}\rightarrow p_{\chi}\right)\right]\frac{2\pi}{V}\delta\left(\varepsilon_{\mathbf{p}_{\chi}}-\varepsilon_{\mathbf{p}{}_{\chi}}\right) \nonumber \\[10pt]={} &-\sum_{\mathrm{partial}}\int\mathrm{d}^{4}z\,\mathrm{d}^{4}z'\,\mathrm{d}^{4}y\,\mathrm{d}^{4}y'\cdots\frac{e^{-i\left(\mathbf{p}{}_{\chi}\cdot\mathbf{z}-\varepsilon{}_{\mathbf{p}_{\chi}}t_{z}\right)}}{\sqrt{V}}G\left(z,\underline{z}';y,\underline{y}',\cdots\right)\frac{e^{i\left(\mathbf{p}{}_{\chi}\cdot\mathbf{z'}-\varepsilon{}_{\mathbf{p}_{\chi}}t_{z'}\right)}}{\sqrt{V}}\nonumber \\[10pt]
={} &
-\sum\left(
\vcenter{\hbox{\tikzset{every picture/.style={line width=0.75pt}}
\begin{tikzpicture}[scale=0.8,
  thick,
  DMfermion/.style={draw=gray!70,
    postaction={decorate},
    decoration={
      markings,
      mark=at position 0.57 with {\arrow{Triangle[scale=0.65, reversed]}}
    },
    thick
  },
  fermion/.style={
    postaction={decorate},
    decoration={
      markings,
      mark=at position 0.57 with {\arrow{Triangle[scale=0.65, reversed]}}
    },
    thick
  },
  photon/.style={
    dashed,          
    thick
  },
fermionloop/.style={
    postaction={decorate},
    decoration={
      markings,
      mark=at position 0.79 with {\arrow{Triangle[scale=0.65]}},
      },
    thick
  },
  momentum/.style={
    postaction={decorate},
    decoration={
      markings,
      mark=at position 0.28 with {\arrow{Triangle[scale=0.65]}}
    }
  }
]
\coordinate (A) at (-2,0);
\coordinate (B) at (-1,0);
\coordinate (C) at (1,0);
\coordinate (D) at (2,0);
\coordinate (X) at (0,0.9); 
\draw[DMfermion,line width=1.2pt] (A) node[] {$$} -- (B) node[midway,above=2pt,black] {$\mathbf{p}_{\chi}$} node[below=5pt] {$$};
\draw[DMfermion,line width=1.2pt] (B) node[] {$$} -- (C) node[midway,below] {$$} node[right] {};
\draw[DMfermion,line width=1.2pt] (C) node[] {$$} -- (D) node[midway,above=2pt,black] {$\mathbf{p}_{\chi}$}node[midway,below] {$$} node[right] {};
\filldraw[fill=black] (-1,0) circle[radius=1.2pt];
\filldraw[fill=black] (1,0) circle[radius=1.2pt];
\draw[photon,line width=1.2pt] (B) to[out=90,in=180] coordinate[pos=0.7] (X1) (X);
\draw[photon,line width=1.2pt] (X) to[out=0,in=90] coordinate[pos=0.7] (X2) (C);
\filldraw[lightgray!50] (X) circle[radius=0.5cm]; 
\draw[fermionloop,line width=1.2pt] (X)  circle[radius=0.5cm];
\path[momentum] (X)  circle[radius=0.5cm];
\filldraw[fill=black] (-0.49,0.80)node[left=5pt] {$$} circle[radius=1.2pt];
\filldraw[fill=black] (0.49,0.80) node[right=5pt] {$$} circle[radius=1.2pt];
\draw[line width=1.2pt] (C) circle[radius=0.12];
\draw[line width=1.2pt] (0.49,0.80) circle[radius=0.12];
\end{tikzpicture}}}
\right),
\label{eq:G_G_diagram}
\end{align}
where the sum on the RHS runs over all partially circled diagrams,
while on the LHS $i\mathcal{M}-i\mathcal{M}^{*}=-2\,\mathrm{Im}\mathcal{M}$.
Since two ends of the massive mediator propagator must be circled
in pairs, the circled and non-circled terms associated with the diagram
\begin{tikzpicture}[baseline={(0,0.1)},scale=0.5,
  thick,
  DMfermion/.style={draw=gray!70,
    postaction={decorate},
    decoration={
      markings,
      mark=at position 0.52 with {\arrow{Triangle[scale=0.65, reversed]}}
    },
    thick
  },
  photon/.style={
    dashed,          
    thick
  }]

\coordinate (A) at (-2,0);
\coordinate (B) at (-1,0);
\coordinate (C) at (1,0);
\coordinate (D) at (2,0);
\coordinate (X) at (0,0.8); 

\draw[DMfermion,line width=1.2pt] (B) -- (C) node[midway,below] {$$} node[right] {};
\draw[DMfermion,line width=1.2pt] (A) -- (B) node[midway,below] {$$} node[right] {};
\draw[DMfermion,line width=1.2pt] (C) -- (D) node[midway,below] {$$} node[right] {};
\filldraw[fill=black] (B) circle[radius=1.2pt];
\filldraw[fill=black] (C) circle[radius=1.2pt];

\draw[photon,line width=1.2pt] (B) to[out=90,in=180] coordinate[pos=0.7] (X1) (X);
\draw[photon,line width=1.2pt] (X) to[out=0,in=90] coordinate[pos=0.7] (X2) (C);
\node at(B)[] {$$};
\node at(C)[] {$$};
\end{tikzpicture} in Eq.~(\ref{eq:G_function1}) is not included on the RHS in Eq.~(\ref{eq:G_G_diagram}).
To see what we have achieved so far we consider the simplest case
where \begin{tikzpicture}[baseline={(0,0.4)},scale=0.6,
  thick,
fermionloop/.style={
    postaction={decorate},
    decoration={
      markings},
    thick},
momentum/.style={
    postaction={decorate},
    decoration={
      markings,
      mark=at position 0.28 with {\arrow{Triangle[scale=0.65]}},
      mark=at position 0.8 with {\arrow{Triangle[scale=0.65]}},
    }
  }]
\coordinate (X) at (0,0.9); 
\filldraw[lightgray!50] (X) circle[radius=0.5cm]; 
\draw[fermionloop,line width=1.2pt] (X)  circle[radius=0.5cm];
\path[momentum] (X)  circle[radius=0.5cm];
\filldraw[fill=black] (-0.5,0.84)node[left=1.5pt] {\small$$} circle[radius=1.2pt];
\filldraw[fill=black] (0.5,0.84) node[right=1.5pt] {\small$$} circle[radius=1.2pt];
\end{tikzpicture}$=$
\begin{tikzpicture}[baseline={(0,0.4)},scale=0.6,
  thick,
fermionloop/.style={
    postaction={decorate},
    decoration={
      markings},
    thick},
momentum/.style={
    postaction={decorate},
    decoration={
      markings,
      mark=at position 0.28 with {\arrow{Triangle[scale=0.65]}},
      mark=at position 0.8 with {\arrow{Triangle[scale=0.65]}},
    }
  }]
\coordinate (X) at (0,0.9); 
\draw[fermionloop,line width=1.2pt] (X)  circle[radius=0.5cm];
\path[momentum] (X)  circle[radius=0.5cm];
\filldraw[fill=black] (-0.5,0.84)node[left=1.5pt] {\small$$} circle[radius=1.2pt];
\filldraw[fill=black] (0.5,0.84) node[right=1.5pt] {\small$$} circle[radius=1.2pt];
\end{tikzpicture},
and combining Eq.~(\ref{eq:decay_rate}), Eq.~(\ref{eq:G_G_diagram})
and Feynman rules in Eqs.~(\ref{eq:fermion3*}-\ref{eq:dark_photon*})
leads to \begin{align}
& 2~\mathrm{Im}\left(
\begin{tikzpicture}[baseline={(0,0.4)}, scale=0.8,       
  thick,
  DMfermion/.style={draw=gray!70,
    postaction={decorate},
    decoration={
      markings,
      mark=at position 0.57 with {\arrow{Triangle[scale=0.65, reversed]}}
    },
    thick
  },
photon/.style={
    dashed,          
    thick
  }]
\coordinate (A) at (-2,0);
\coordinate (B) at (-1,0);
\coordinate (C) at (1,0);
\coordinate (D) at (2,0);
\coordinate (X) at (0,0.9); 
\draw[DMfermion,line width=1.2pt] (A) node[] {$$} -- (B) node[midway,above=2pt,black] {$\mathbf{p}_{\chi}$} node[below=5pt] {$$};
\draw[DMfermion,line width=1.2pt] (B) node[] {$$} -- (C) node[midway,below] {$$} node[right] {};
\draw[DMfermion,line width=1.2pt] (C) node[] {$$} -- (D) node[midway,above=2pt,black] {$\mathbf{p}_{\chi}$}node[midway,below] {$$} node[right] {};
\filldraw[fill=black] (-1,0) circle[radius=1.2pt];
\filldraw[fill=black] (1,0) circle[radius=1.2pt];
\draw[photon,line width=1.2pt] (B) to[out=90,in=180] coordinate[pos=0.7] (X1) (X);
\draw[photon,line width=1.2pt] (X) to[out=0,in=90] coordinate[pos=0.7] (X2) (C);
\end{tikzpicture}+
\begin{tikzpicture}[baseline={(0,0.4)}, scale=0.8,
  thick,
  DMfermion/.style={draw=gray!70,
    postaction={decorate},
    decoration={
      markings,
      mark=at position 0.57 with {\arrow{Triangle[scale=0.65, reversed]}}
    },
    thick
  },
  fermion/.style={
    postaction={decorate},
    decoration={
      markings,
      mark=at position 0.57 with {\arrow{Triangle[scale=0.65, reversed]}}
    },
    thick
  },
  photon/.style={
    dashed,          
    thick
  },
fermionloop/.style={
    postaction={decorate},
    decoration={
      markings,
      mark=at position 0.79 with {\arrow{Triangle[scale=0.65]}},
      },
    thick
  },
  momentum/.style={
    postaction={decorate},
    decoration={
      markings,
      mark=at position 0.28 with {\arrow{Triangle[scale=0.65]}}
    }
  }
]
\coordinate (A) at (-2,0);
\coordinate (B) at (-1,0);
\coordinate (C) at (1,0);
\coordinate (D) at (2,0);
\coordinate (X) at (0,0.9); 
\draw[DMfermion,line width=1.2pt] (A) node[] {$$} -- (B) node[midway,above=2pt,black] {$\mathbf{p}_{\chi}$} node[below=5pt] {$$};
\draw[DMfermion,line width=1.2pt] (B) node[] {$$} -- (C) node[midway,below] {$$} node[right] {};
\draw[DMfermion,line width=1.2pt] (C) node[] {$$} -- (D) node[midway,above=2pt,black] {$\mathbf{p}_{\chi}$}node[midway,below] {$$} node[right] {};
\filldraw[fill=black] (-1,0) circle[radius=1.2pt];
\filldraw[fill=black] (1,0) circle[radius=1.2pt];
\draw[photon,line width=1.2pt] (B) to[out=90,in=180] coordinate[pos=0.7] (X1) (X);
\draw[photon,line width=1.2pt] (X) to[out=0,in=90] coordinate[pos=0.7] (X2) (C);
\filldraw[white] (X) circle[radius=0.5cm]; 
\draw[fermionloop,line width=1.2pt] (X)  circle[radius=0.5cm];
\path[momentum] (X)  circle[radius=0.5cm];
\filldraw[fill=black] (-0.49,0.80)node[left=5pt] {$$} circle[radius=1.2pt];
\filldraw[fill=black] (0.49,0.80) node[right=5pt] {$$} circle[radius=1.2pt];
\%draw[line width=1.2pt] (-0.49,0.80) circle[radius=0.12];
\end{tikzpicture}
\right)\nonumber \\
={} & 
\left(
\begin{tikzpicture}[baseline={(0,0.4)}, scale=0.8,
  thick,
  DMfermion/.style={draw=gray!70,
    postaction={decorate},
    decoration={
      markings,
      mark=at position 0.57 with {\arrow{Triangle[scale=0.65, reversed]}}
    },
    thick
  },
  fermion/.style={
    postaction={decorate},
    decoration={
      markings,
      mark=at position 0.57 with {\arrow{Triangle[scale=0.65, reversed]}}
    },
    thick
  },
  photon/.style={
    dashed,          
    thick
  },
fermionloop/.style={
    postaction={decorate},
    decoration={
      markings,
      mark=at position 0.79 with {\arrow{Triangle[scale=0.65]}},
      },
    thick
  },
  momentum/.style={
    postaction={decorate},
    decoration={
      markings,
      mark=at position 0.28 with {\arrow{Triangle[scale=0.65]}}
    }
  }
]
\coordinate (A) at (-2,0);
\coordinate (B) at (-1,0);
\coordinate (C) at (1,0);
\coordinate (D) at (2,0);
\coordinate (X) at (0,0.9); 
\draw[DMfermion,line width=1.2pt] (A) node[] {$$} -- (B) node[midway,above=2pt,black] {$\mathbf{p}_{\chi}$} node[below=5pt] {$$};
\draw[DMfermion,line width=1.2pt] (B) node[] {$$} -- (C) node[midway,below] {$$} node[right] {};
\draw[DMfermion,line width=1.2pt] (C) node[] {$$} -- (D) node[midway,above=2pt,black] {$\mathbf{p}_{\chi}$}node[midway,below] {$$} node[right] {};
\filldraw[fill=black] (-1,0) circle[radius=1.2pt];
\filldraw[fill=black] (1,0) circle[radius=1.2pt];
\draw[photon,line width=1.2pt] (B) to[out=90,in=180] coordinate[pos=0.7] (X1) (X);
\draw[photon,line width=1.2pt] (X) to[out=0,in=90] coordinate[pos=0.7] (X2) (C);
\filldraw[white] (X) circle[radius=0.5cm]; 
\draw[fermionloop,line width=1.2pt] (X)  circle[radius=0.5cm];
\path[momentum] (X)  circle[radius=0.5cm];
\filldraw[fill=black] (-0.49,0.80)node[left=5pt] {$$} circle[radius=1.2pt];
\filldraw[fill=black] (0.49,0.80) node[right=5pt] {$$} circle[radius=1.2pt];
\draw[line width=1.2pt] (C) circle[radius=0.12];
\draw[line width=1.2pt] (0.49,0.80) circle[radius=0.12];
\end{tikzpicture}+
\begin{tikzpicture}[baseline={(0,0.4)}, scale=0.8,
  thick,
  DMfermion/.style={draw=gray!70,
    postaction={decorate},
    decoration={
      markings,
      mark=at position 0.57 with {\arrow{Triangle[scale=0.65, reversed]}}
    },
    thick
  },
  fermion/.style={
    postaction={decorate},
    decoration={
      markings,
      mark=at position 0.57 with {\arrow{Triangle[scale=0.65, reversed]}}
    },
    thick
  },
  photon/.style={
    dashed,          
    thick
  },
fermionloop/.style={
    postaction={decorate},
    decoration={
      markings,
      mark=at position 0.79 with {\arrow{Triangle[scale=0.65]}},
      },
    thick
  },
  momentum/.style={
    postaction={decorate},
    decoration={
      markings,
      mark=at position 0.28 with {\arrow{Triangle[scale=0.65]}}
    }
  }
]
\coordinate (A) at (-2,0);
\coordinate (B) at (-1,0);
\coordinate (C) at (1,0);
\coordinate (D) at (2,0);
\coordinate (X) at (0,0.9); 
\draw[DMfermion,line width=1.2pt] (A) node[] {$$} -- (B) node[midway,above=2pt,black] {$\mathbf{p}_{\chi}$} node[below=5pt] {$$};
\draw[DMfermion,line width=1.2pt] (B) node[] {$$} -- (C) node[midway,below] {$$} node[right] {};
\draw[DMfermion,line width=1.2pt] (C) node[] {$$} -- (D) node[midway,above=2pt,black] {$\mathbf{p}_{\chi}$}node[midway,below] {$$} node[right] {};
\filldraw[fill=black] (-1,0) circle[radius=1.2pt];
\filldraw[fill=black] (1,0) circle[radius=1.2pt];
\draw[photon,line width=1.2pt] (B) to[out=90,in=180] coordinate[pos=0.7] (X1) (X);
\draw[photon,line width=1.2pt] (X) to[out=0,in=90] coordinate[pos=0.7] (X2) (C);
\filldraw[white] (X) circle[radius=0.5cm]; 
\draw[fermionloop,line width=1.2pt] (X)  circle[radius=0.5cm];
\path[momentum] (X)  circle[radius=0.5cm];
\filldraw[fill=black] (-0.49,0.80)node[left=5pt] {$$} circle[radius=1.2pt];
\filldraw[fill=black] (0.49,0.80) node[right=5pt] {$$} circle[radius=1.2pt];
\draw[line width=1.2pt] (B) circle[radius=0.12];
\draw[line width=1.2pt] (-0.49,0.80) circle[radius=0.12];
\end{tikzpicture}
\right)\nonumber \\[10pt]
={} & 
\sum_{\mathbf{p}'_{\chi}}\sum_{i>F}\sum_{j\leq F}~\left|
\vcenter{\hbox{\tikzset{every picture/.style={line width=0.75pt}}
\begin{tikzpicture}[scale=0.8,
  thick,
  DMfermion/.style={draw=gray!70,
    postaction={decorate},
    decoration={
      markings,
      mark=at position 0.57 with {\arrow{Triangle[scale=0.65,reversed]}}
    },
    thick
  },
   fermion/.style={
    postaction={decorate},
    decoration={
      markings,
      mark=at position 0.57 with {\arrow{Triangle[scale=0.65]}}
    },
    thick
  },
  photon/.style={
    dashed,          
    thick
  }]
\coordinate (A) at (-1.1,-0.75);
\coordinate (B) at (-0.65,0);
\coordinate (C) at (-1.1,0.75);
\coordinate (D) at (0.65,0);
\coordinate (E) at (1.1,0.75);
\coordinate (F) at (1.1,-0.75);
\coordinate (H) at (0.85,0);
\coordinate (I) at (1.3,-0.75);
\draw[DMfermion,line width=1.2pt] (B) node[anchor=south east, xshift=-8pt, yshift=5pt] {$\mathbf{p'}_{\chi}$} -- (A) node[midway,below] {};
\draw[DMfermion,line width=1.2pt] (C) node[above] {$$} -- (B) node[anchor=south east, xshift=-10pt, yshift=-23pt] {$\mathbf{p}_{\chi}$};
\draw[photon,line width=1.2pt] (B)  -- (D) node[midway,below] {$$} node[right] {};
\draw[fermion,line width=1.2pt] (D)node[anchor=north west, xshift=12pt, yshift=25pt] {$i$} -- (E) node[midway,below] {$$} node[right] {};
\draw[fermion,line width=1.2pt] (F) -- (D) node[anchor=north west, xshift=12pt, yshift=-7pt] {$j$};
\draw[line width=1.2pt,
    dashed, color=gray,
    decoration={markings, mark=at position 0.57 with {\arrow{Triangle[scale=0.65]}}},
    postaction={decorate}
  ] 
  (H) -- (I);
\filldraw[fill=black] (B) circle[radius=1.2pt];
\filldraw[fill=black] (D) circle[radius=1.2pt];
\end{tikzpicture}}}\right|^{2},
\label{eq:cutting_diagram}
\end{align}where the first term on the second line vanishes because of the Feynman
rule in Eq.~(\ref{eq:DM_0}). The key point in the above analysis
is that we split the cut propagator \begin{tikzpicture}[baseline={(0,-0.05)}, scale=1]
\draw[line width=1.2pt,
    decoration={markings, mark=at position 0.57 with {\arrow{Triangle[scale=0.65]}}},
    postaction={decorate}
  ] 
  (0.5,0) -- (-0.5,0) ;
  \filldraw[black] (-0.5,0) circle[radius=1.2pt]
    node[left=2pt] {$x$};
  \filldraw[black] (0.5,0) circle[radius=1.2pt]
    node[right=2pt] {$y$};
\draw[line width=1.2pt] (-0.5,0) circle[radius=0.1];
\end{tikzpicture}  into $u_{k}\left(\mathbf{x}\right)e^{-i\varepsilon_{k}t_{x}}$ and
$\left(u_{k}\left(\mathbf{y}\right)e^{-i\varepsilon_{k}t_{y}}\right)^{*}$:
the latter amounts to the outgoing external leg while the former corresponds
to its complex conjugate. The act of circling only one end of a fermion
line can also be conceived as cutting that propagator. The cut divides
the second diagram on the second line symmetrically into two halves.
Therefore, Eq.~(\ref{eq:cutting_diagram}) is exactly the cutting
rule and optical theorem at the one-loop level. It should also be
noted that, since the discussion is carried out within the framework
of one-shot perturbation theory, field renormalization constants $Z_{e}$,
as well as corrections to the external lines of electron-hole quasi-particles
in the LSZ formula, are irrelevant at this stage. 

The relation in Eq.~(\ref{eq:G_G_diagram}) can be generalized to
any flip-invariant diagrams, up to the most general case, where on
the RHS the gray blob is replaced by the electron blob \blobFigure[baseline={(0,0.4)},scale=0.6]
in Eq.~(\ref{eq:G_function1}). For a diagram violating the symmetry,
adding its symmetrically complementary partner can restore invariance,
which is a natural requirement for physically valid self-energy expression.
Therefore one can only consider flip-symmetric diagrams in a diagram-by-diagram
cutting-rule analysis. 

In the following, we will illustrate how to implement the above cutting
rules in two specific examples.

\subsection{vertex correction}

In the first example we apply the derived cutting rule to a more complex
example, where \begin{tikzpicture}[baseline={(0,0.4)},scale=0.6,
  thick,
fermionloop/.style={
    postaction={decorate},
    decoration={
      markings},
    thick},
momentum/.style={
    postaction={decorate},
    decoration={
      markings,
      mark=at position 0.28 with {\arrow{Triangle[scale=0.65]}},
      mark=at position 0.8 with {\arrow{Triangle[scale=0.65]}},
    }
  }]
\coordinate (X) at (0,0.9); 
\filldraw[lightgray!50] (X) circle[radius=0.5cm]; 
\draw[fermionloop,line width=1.2pt] (X)  circle[radius=0.5cm];
\path[momentum] (X)  circle[radius=0.5cm];
\filldraw[fill=black] (-0.5,0.84)node[left=1.5pt] {\small$$} circle[radius=1.2pt];
\filldraw[fill=black] (0.5,0.84) node[right=1.5pt] {\small$$} circle[radius=1.2pt];
\end{tikzpicture}$=$
\begin{tikzpicture}[baseline={(0,0.4)},scale=0.6,
  thick,
fermionloop/.style={
    postaction={decorate},
    decoration={
      markings},
    thick},
photon1/.style={
    thick,
    decorate,
    decoration={
      snake,
      amplitude=1.5pt,
      segment length=4pt
    }
  },
momentum/.style={
    postaction={decorate},
    decoration={
      markings,
      mark=at position 0.28 with {\arrow{Triangle[scale=0.65]}},
      mark=at position 0.8 with {\arrow{Triangle[scale=0.65]}},
    }
  }]
\coordinate (X) at (0,0.9); 
\draw (X) circle[radius=0.5cm];
\draw[fermionloop,line width=1.2pt] (X)  circle[radius=0.5cm];
\path[momentum] (X)  circle[radius=0.5cm];
\filldraw[fill=black] (-0.5,0.84)node[left=1.5pt] {\small$$} circle[radius=1.2pt];
\filldraw[fill=black] (0.5,0.84) node[right=1.5pt] {\small$$} circle[radius=1.2pt];
\end{tikzpicture}$+$ 
\begin{tikzpicture}[baseline={(0,0.4)},scale=0.6,
  thick,
fermionloop/.style={
    postaction={decorate},
    decoration={
      markings},
    thick},
photon1/.style={
    thick,
    decorate,
    decoration={
      snake,
      amplitude=1.5pt,
      segment length=4pt
    }
  },
momentum/.style={
    postaction={decorate},
    decoration={
      markings,
      mark=at position 0.35 with {\arrow{Triangle[scale=0.65]}},
      mark=at position 0.85 with {\arrow{Triangle[scale=0.65]}},
     }
  }]
\coordinate (X) at (0,0.9); 
\coordinate (X1) at (0,1.4);
\coordinate (X2) at (0,0.4);
\draw (X) circle[radius=0.5cm];
\draw[fermionloop,line width=1.2pt] (X)  circle[radius=0.5cm];
\draw[photon1,line width=1.2pt] (X1) -- (X2);
\path[momentum] (X)  circle[radius=0.5cm];
\filldraw[fill=black] (-0.5,0.84)node[left=1.5pt] {\small$$} circle[radius=1.2pt];
\filldraw[fill=black] (0.5,0.84) node[right=1.5pt] {\small$$} circle[radius=1.2pt];
\end{tikzpicture}.
 The cutting operation can be implemented as follows, 
\tikzset{
  DMfermion/.style={
    draw=gray!70,
    postaction={decorate},
    decoration={
      markings,
      mark=at position 0.57 with {\arrow{Triangle[scale=0.65, reversed]}}
    },
    thick
  },
  fermion/.style={
    postaction={decorate},
    decoration={
      markings,
      mark=at position 0.57 with {\arrow{Triangle[scale=0.65]}}
    },
    thick
  },
  photon/.style={
    dashed, thick
  },
  fermionloop/.style={
    postaction={decorate},
    decoration={
      markings,
      mark=at position 0.84 with {\arrow{Triangle[scale=0.65]}}
    },
    thick
  },
  photon1/.style={
    thick,
    decorate,
    decoration={
      snake,
      amplitude=2pt,
      segment length=5pt
    }
  },
  momentum/.style={
    postaction={decorate},
    decoration={
      markings,
      mark=at position 0.34 with {\arrow{Triangle[scale=0.65]}}
    }
  }
}

\newcommand{\commoncoords}{
  \coordinate (A) at (-2,0);
  \coordinate (B) at (-1,0);
  \coordinate (C) at (1,0);
  \coordinate (D) at (2,0);
  \coordinate (X) at (0,0.9);
}

\newcommand{\fermionloopstructure}{
  \filldraw[white] (X) circle[radius=0.5cm];
  \draw[fermionloop,line width=1.2pt] (X) circle[radius=0.5cm];
  \path[momentum] (X) circle[radius=0.5cm];
  \filldraw[fill=black] (-0.49,0.8) circle[radius=1.2pt];
  \filldraw[fill=black] (0.49,0.8) circle[radius=1.2pt];
}

\newcommand{\mainstructure}[1]{
  \commoncoords
  \draw[DMfermion,line width=1.2pt] (A) -- (B) node[midway,above=2pt] {$\mathbf{p}_{\chi}$};
  \draw[DMfermion,line width=1.2pt] (B) -- (C);
  \draw[DMfermion,line width=1.2pt] (C) -- (D) node[midway,above=2pt] {$\mathbf{p}_{\chi}$};
  \filldraw[fill=black] (-1,0) circle[radius=1.2pt];
  \filldraw[fill=black] (1,0) circle[radius=1.2pt];
  \draw[photon,line width=1.2pt] (B) to[out=90,in=180] (X);
  \draw[photon,line width=1.2pt] (X) to[out=0,in=90] (C);
  #1
}
\begin{align}
& 2~\mathrm{Im} \left(
  \begin{tikzpicture}[baseline={(0,0.4)}, scale=0.8, thick]
    \mainstructure{}
  \end{tikzpicture}
  +
  \begin{tikzpicture}[baseline={(0,0.4)}, scale=0.8, thick]
    \mainstructure{\fermionloopstructure}
  \end{tikzpicture}
  +
  \begin{tikzpicture}[baseline={(0,0.4)}, scale=0.8, thick]
    \mainstructure{
      \fermionloopstructure
      \coordinate (Z1) at (0,1.4);
      \coordinate (Z2) at (0,0.4);
      \draw[photon1,line width=1.2pt] (Z1) -- (Z2);
    }
  \end{tikzpicture}
\right) \nonumber \\
={} & 
  \begin{tikzpicture}[baseline={(0,0.4)}, scale=0.8, thick]
    \mainstructure{
      \fermionloopstructure
      \draw[gray!90,dotted,line width=1.5pt] ($(X)+(0,-1.5)$) -- ($(X)+(0,1.2)$);
    }
  \end{tikzpicture}
  +
  \begin{tikzpicture}[baseline={(0,0.4)}, scale=0.8, thick]
    \mainstructure{
      \fermionloopstructure
      \draw[photon1,line width=1.2pt] (0,1.4) -- (0,0.4);
      \draw[gray!90,dotted,line width=1.5pt] 
        ($(X)+(-0.5,-1.5)$) to[out=70, in=-70] ($(X)+(-0.5,1.2)$);
    }
  \end{tikzpicture}
  +
  \begin{tikzpicture}[baseline={(0,0.4)}, scale=0.8, thick]
    \mainstructure{
      \fermionloopstructure
      \draw[photon1,line width=1.2pt] (0,1.4) -- (0,0.4);
      \draw[gray!90,dotted,line width=1.5pt] 
        ($(X)+(0.5,-1.5)$) to[out=110, in=-110] ($(X)+(0.5,1.2)$);
    }
  \end{tikzpicture}
  \nonumber \\[10pt]
={} & 
\sum_{\mathbf{p}'_{\chi}}\sum_{i>F}\sum_{j\leq F}~\left[~\left|
\begin{tikzpicture}[baseline={(0,0)}, scale=0.8, thick]
\coordinate (A) at (-1.1,-0.75);
\coordinate (B) at (-0.65,0);
\coordinate (C) at (-1.1,0.75);
\coordinate (D) at (0.65,0);
\coordinate (E) at (1.1,0.75);
\coordinate (F) at (1.1,-0.75);
\coordinate (H) at (0.85,0);
\coordinate (I) at (1.3,-0.75);
\draw[DMfermion,line width=1.2pt] (B) node[anchor=south east, xshift=-8pt, yshift=5pt] {$\mathbf{p'}_{\chi}$} -- (A);
\draw[DMfermion,line width=1.2pt] (C) -- (B) node[anchor=south east, xshift=-10pt, yshift=-23pt] {$\mathbf{p}_{\chi}$};
\draw[photon,line width=1.2pt] (B) -- (D);
\draw[fermion,line width=1.2pt] (D) node[anchor=north west, xshift=12pt, yshift=25pt] {$i$} -- (E);
\draw[fermion,line width=1.2pt] (F) -- (D) node[anchor=north west, xshift=12pt, yshift=-7pt] {$j$};
\filldraw[fill=black] (B) circle[radius=1.2pt];
\filldraw[fill=black] (D) circle[radius=1.2pt];
\end{tikzpicture}\right|^{2}
+
\left(\begin{tikzpicture}[baseline={(0,0)}, scale=0.8, thick]
\coordinate (A) at (-1.1,-0.75);
\coordinate (B) at (-0.65,0);
\coordinate (C) at (-1.1,0.75);
\coordinate (D) at (0.65,0);
\coordinate (E) at (1.1,0.75);
\coordinate (F) at (1.1,-0.75);
\draw[DMfermion,line width=1.2pt] (B) node[anchor=south east, xshift=-8pt, yshift=5pt] {$\mathbf{p'}_{\chi}$} -- (A);
\draw[DMfermion,line width=1.2pt] (C) -- (B) node[anchor=south east, xshift=-10pt, yshift=-23pt] {$\mathbf{p}_{\chi}$};
\draw[photon,line width=1.2pt] (B) -- (D);
\draw[fermion,line width=1.2pt] (D) node[anchor=north west, xshift=12pt, yshift=25pt] {$i$} -- (E);
\draw[fermion,line width=1.2pt] (F) -- (D) node[anchor=north west, xshift=12pt, yshift=-7pt] {$j$};
\filldraw[fill=black] (B) circle[radius=1.2pt];
\filldraw[fill=black] (D) circle[radius=1.2pt];
\end{tikzpicture}\right)^{*}
\left(\begin{tikzpicture}[baseline={(0,0)}, scale=0.8, thick]
\coordinate (A) at (-1.1,-0.75);
\coordinate (B) at (-0.65,0);
\coordinate (C) at (-1.1,0.75);
\coordinate (D) at (0.65,0);
\coordinate (E) at (1.1,0.75);
\coordinate (F) at (1.1,-0.75);
\draw[DMfermion,line width=1.2pt] (B) node[anchor=south east, xshift=-8pt, yshift=5pt] {$\mathbf{p'}_{\chi}$} -- (A);
\draw[DMfermion,line width=1.2pt] (C) -- (B) node[anchor=south east, xshift=-10pt, yshift=-23pt] {$\mathbf{p}_{\chi}$};
\draw[photon,line width=1.2pt] (B) -- (D);
\draw[fermion,line width=1.2pt] (D) node[anchor=north west, xshift=12pt, yshift=25pt] {$i$} -- (E);
\draw[fermion,line width=1.2pt] (F) -- (D) node[anchor=north west, xshift=12pt, yshift=-7pt] {$j$};
\filldraw[fill=black] (B) circle[radius=1.2pt];
\filldraw[fill=black] (D) circle[radius=1.2pt];
\draw[photon1,line width=1.2pt] (0.95,0.55) to[out=-40, in=40] (0.95,-0.55);
\end{tikzpicture}\right)
\right. \nonumber \\
{} &+ \left.\left(\begin{tikzpicture}[baseline={(0,0)}, scale=0.8, thick]
\coordinate (A) at (-1.1,-0.75);
\coordinate (B) at (-0.65,0);
\coordinate (C) at (-1.1,0.75);
\coordinate (D) at (0.65,0);
\coordinate (E) at (1.1,0.75);
\coordinate (F) at (1.1,-0.75);
\draw[DMfermion,line width=1.2pt] (B) node[anchor=south east, xshift=-8pt, yshift=5pt] {$\mathbf{p'}_{\chi}$} -- (A);
\draw[DMfermion,line width=1.2pt] (C) -- (B) node[anchor=south east, xshift=-10pt, yshift=-23pt] {$\mathbf{p}_{\chi}$};
\draw[photon,line width=1.2pt] (B) -- (D);
\draw[fermion,line width=1.2pt] (D) node[anchor=north west, xshift=12pt, yshift=25pt] {$i$} -- (E);
\draw[fermion,line width=1.2pt] (F) -- (D) node[anchor=north west, xshift=12pt, yshift=-7pt] {$j$};
\filldraw[fill=black] (B) circle[radius=1.2pt];
\filldraw[fill=black] (D) circle[radius=1.2pt];
\draw[photon1,line width=1.2pt] (0.95,0.55) to[out=-40, in=40] (0.95,-0.55);
\end{tikzpicture}\right)^{*}
\left(\begin{tikzpicture}[baseline={(0,0)}, scale=0.8, thick]
\coordinate (A) at (-1.1,-0.75);
\coordinate (B) at (-0.65,0);
\coordinate (C) at (-1.1,0.75);
\coordinate (D) at (0.65,0);
\coordinate (E) at (1.1,0.75);
\coordinate (F) at (1.1,-0.75);
\draw[DMfermion,line width=1.2pt] (B) node[anchor=south east, xshift=-8pt, yshift=5pt] {$\mathbf{p'}_{\chi}$} -- (A);
\draw[DMfermion,line width=1.2pt] (C) -- (B) node[anchor=south east, xshift=-10pt, yshift=-23pt] {$\mathbf{p}_{\chi}$};
\draw[photon,line width=1.2pt] (B) -- (D);
\draw[fermion,line width=1.2pt] (D) node[anchor=north west, xshift=12pt, yshift=25pt] {$i$} -- (E);
\draw[fermion,line width=1.2pt] (F) -- (D) node[anchor=north west, xshift=12pt, yshift=-7pt] {$j$};
\filldraw[fill=black] (B) circle[radius=1.2pt];
\filldraw[fill=black] (D) circle[radius=1.2pt];
\end{tikzpicture}\right)\right],
\label{eq:cutting_diagram3}
\end{align} which means that the higher order loop term \begin{tikzpicture}[baseline={(0,0.4)},scale=0.6,
  thick,
fermionloop/.style={
    postaction={decorate},
    decoration={
      markings},
    thick},
photon1/.style={
    thick,
    decorate,
    decoration={
      snake,
      amplitude=1.5pt,
      segment length=4pt
    }
  },
momentum/.style={
    postaction={decorate},
    decoration={
      markings,
      mark=at position 0.35 with {\arrow{Triangle[scale=0.65]}},
      mark=at position 0.85 with {\arrow{Triangle[scale=0.65]}},
     }
  }]
\coordinate (X) at (0,0.9); 
\coordinate (X1) at (0,1.4);
\coordinate (X2) at (0,0.4);
\draw (X) circle[radius=0.5cm];
\draw[fermionloop,line width=1.2pt] (X)  circle[radius=0.5cm];
\draw[photon1,line width=1.2pt] (X1) -- (X2);
\path[momentum] (X)  circle[radius=0.5cm];
\filldraw[fill=black] (-0.5,0.84)node[left=1.5pt] {\small$$} circle[radius=1.2pt];
\filldraw[fill=black] (0.5,0.84) node[right=1.5pt] {\small$$} circle[radius=1.2pt];
\end{tikzpicture} gives rise to an electron vertex correction term \begin{tikzpicture}[baseline={(0,-0.4)},scale=0.65]
  \draw[line width=1.2pt,
    decoration={
      markings,
      mark=at position 0.35 with {\arrow{Triangle[scale=0.65]}},
      mark=at position 0.75 with {\arrow{Triangle[scale=0.65]}}
    },
    postaction={decorate}
  ] (0,0) arc [start angle=90, end angle=270, radius=0.5cm];
 \draw[
    decoration={snake, amplitude=0.5mm, segment length=1.2mm}, 
    decorate,
    line width=1.2pt
  ]
  (-0.3,-0.1) to[out=-40, in=40] (-0.3,-0.9);
\filldraw[fill=black] (-0.5,-0.5)node[left=1.5pt] {\small$$} circle[radius=1.3pt];
\end{tikzpicture}~
in the DM decay calculation. 

\subsection{screening in the medium}

Now we consider an example of practical significance, which corresponds
to the situation where one uses a fermion loop to represent the 1PI
in Eq.~(\ref{eq:blob_series}), \emph{i.e}., \PI[baseline=-0.75ex,scale=0.7]{0.5}$\simeq$
\begin{tikzpicture}[baseline={(0,0.4)},scale=0.6,
  thick,
fermionloop/.style={
    postaction={decorate},
    decoration={
      markings},
    thick},
momentum/.style={
    postaction={decorate},
    decoration={
      markings,
      mark=at position 0.28 with {\arrow{Triangle[scale=0.65]}},
      mark=at position 0.8 with {\arrow{Triangle[scale=0.65]}},
    }
  }]
\coordinate (X) at (0,0.9); 
\draw (X) circle[radius=0.5cm];
\draw[fermionloop,line width=1.2pt] (X)  circle[radius=0.5cm];
\path[momentum] (X)  circle[radius=0.5cm];
\filldraw[fill=black] (-0.5,0.84)node[left=1.5pt] {\small$$} circle[radius=1.2pt];
\filldraw[fill=black] (0.5,0.84) node[right=1.5pt] {\small$$} circle[radius=1.2pt];
\end{tikzpicture}, or namely, the \emph{random phase approximation} (RPA). To be specific,
we are investigating the cutting rules for the diagrams \begin{equation}
\begin{tikzpicture}[baseline={(0,0.18)}, scale=1,       
  thick,
  DMfermion/.style={
    draw=gray!70,
    postaction={decorate},
    decoration={
      markings,
      mark=at position 0.57 with {\arrow{Triangle[scale=0.65, reversed]}}
    },
    thick
  },
  photon/.style={
    dashed,
    thick
  }]
  \coordinate (A) at (-1.5,0);
  \coordinate (B) at (-0.8,0);
  \coordinate (C) at (0.8,0);
  \coordinate (D) at (1.5,0);
  \coordinate (X) at (0,0.7); 
  \draw[DMfermion,line width=1.2pt] (A) -- (B) node[midway,above=2pt,black] {\small$\mathbf{p}_{\chi}$};
  \draw[DMfermion,line width=1.2pt] (B) -- (C);
  \draw[DMfermion,line width=1.2pt] (C) -- (D) node[midway,above=2pt,black] {\small$\mathbf{p}_{\chi}$};
  \filldraw[fill=black] (B) circle[radius=1.2pt];
  \filldraw[fill=black] (C) circle[radius=1.2pt];
  \draw[photon,line width=1.2pt] (B) to[out=90,in=180] coordinate[pos=0.7] (X1) (X);
  \draw[photon,line width=1.2pt] (X) to[out=0,in=90] coordinate[pos=0.7] (X2) (C);
\end{tikzpicture}
+
\begin{tikzpicture}[baseline={(0,0.18)}, scale=1,
  thick,
  DMfermion/.style={
    draw=gray!70,
    postaction={decorate},
    decoration={
      markings,
      mark=at position 0.57 with {\arrow{Triangle[scale=0.65, reversed]}}
    },
    thick
  },
  photon/.style={
    dashed,
    thick
  },
  fermionloop/.style={
    postaction={decorate},
    decoration={
      markings,
      mark=at position 0.79 with {\arrow{Triangle[scale=0.65]}}
    },
    thick
  },
  momentum/.style={
    postaction={decorate},
    decoration={
      markings,
      mark=at position 0.28 with {\arrow{Triangle[scale=0.65]}}
    }
  }]
  \coordinate (A) at (-1.5,0);
  \coordinate (B) at (-0.8,0);
  \coordinate (C) at (0.8,0);
  \coordinate (D) at (1.5,0);
  \coordinate (X) at (0,0.7); 
  \draw[DMfermion,line width=1.2pt] (A) -- (B) node[midway,above=2pt,black] {\small$\mathbf{p}_{\chi}$};
  \draw[DMfermion,line width=1.2pt] (B) -- (C);
  \draw[DMfermion,line width=1.2pt] (C) -- (D) node[midway,above=2pt,black] {\small$\mathbf{p}_{\chi}$};
  \filldraw[fill=black] (B) circle[radius=1.2pt];
  \filldraw[fill=black] (C) circle[radius=1.2pt];
  \draw[photon,line width=1.2pt] (B) to[out=90,in=180] coordinate[pos=0.7] (X1) (X);
  \draw[photon,line width=1.2pt] (X) to[out=0,in=90] coordinate[pos=0.7] (X2) (C);
  \filldraw[white] (X) circle[radius=0.30cm]; 
  \draw[fermionloop,line width=1.2pt] (X) circle[radius=0.3cm];
  \path[momentum] (X) circle[radius=0.3cm];
  \filldraw[fill=black] (-0.29,0.68) circle[radius=1.2pt];
  \filldraw[fill=black] (0.29,0.68) circle[radius=1.2pt];
\end{tikzpicture}
+
\begin{tikzpicture}[baseline={(0,0.18)}, scale=1,
  thick,
  DMfermion/.style={
    draw=gray!70,
    postaction={decorate},
    decoration={
      markings,
      mark=at position 0.57 with {\arrow{Triangle[scale=0.65, reversed]}}
    },
    thick
  },
  photon/.style={
    dashed,
    thick
  },
  photon1/.style={
    thick,
    decorate,
    decoration={
      snake,
      amplitude=1.5pt,
      segment length=4pt
    }
  },
  fermionloop/.style={
    postaction={decorate},
    decoration={
      markings,
      mark=at position 0.79 with {\arrow{Triangle[scale=0.65]}}
    },
    thick
  },
  momentum/.style={
    postaction={decorate},
    decoration={
      markings,
      mark=at position 0.28 with {\arrow{Triangle[scale=0.65]}}
    }
  }]
  \coordinate (A) at (-1.8,0);
  \coordinate (B) at (-1,0);
  \coordinate (C) at (1,0);
  \coordinate (D) at (1.8,0);
  \coordinate (X) at (-0.6,0.7);
  \coordinate (X1) at (0.6,0.7);
  \draw[DMfermion,line width=1.2pt] (A) -- (B) node[midway,above=2pt,black] {\small$\mathbf{p}_{\chi}$};
  \draw[DMfermion,line width=1.2pt] (B) -- (C);
  \draw[DMfermion,line width=1.2pt] (C) -- (D) node[midway,above=2pt,black] {\small$\mathbf{p}_{\chi}$};
  \filldraw[fill=black] (B) circle[radius=1.2pt];
  \filldraw[fill=black] (C) circle[radius=1.2pt];
  \draw[photon,line width=1.2pt] (B) to[out=90,in=180] (X);
  \draw[photon1,line width=1.2pt] (X) -- (X1);
  \draw[photon,line width=1.2pt] (C) to[out=90,in=0] (X1);
  \filldraw[white] (X) circle[radius=0.30cm];
  \filldraw[white] (X1) circle[radius=0.30cm];
  \draw[fermionloop,line width=1.2pt] (X) circle[radius=0.3cm];
  \draw[fermionloop,line width=1.2pt] (X1) circle[radius=0.3cm];
  \path[momentum] (X) circle[radius=0.3cm];
  \path[momentum] (X1) circle[radius=0.3cm];
  \filldraw[fill=black] (-0.86,0.55) circle[radius=1.2pt];
  \filldraw[fill=black] (0.86,0.55) circle[radius=1.2pt];
\end{tikzpicture}
+\cdots,
\label{eq:RPA_series}
\end{equation} where the sum of infinite series describes the screening and collective
effects of the polarized electron medium. This particular form of
self-energy is analogous to the $G_{0}W_{0}$ approximation in the
DFT language, which can be heuristically thought of as a dynamic HF
approximation with the bare Coulomb interaction being substituted
by an energy-dependent screened interaction. In this case, $G_{0}$
represents the undressed DM particle Green's function (or equivalently,
propagator); $W_{0}$ corresponds to the DM-electron Yukawa propagator
screened by the KS electron bubbles in the context of RPA, and the
vertex correction is neglected. 

Integrating the underlining rules in Eqs.~(\ref{eq:fermion3*}-\ref{eq:dark_photon*})
with energy conservation, it is not difficult to observe that only
a limited set of partially circled diagrams are relevant in the discussion
parallel to Eq.~(\ref{eq:cutting_diagram}). For instance, for a situation
like the following, we have \begin{equation}
\vcenter{\hbox{\tikzset{every picture/.style={line width=0.75pt}}
\begin{tikzpicture}[
  thick,
  fermion/.style={
    postaction={decorate},
    decoration={
      markings,
      mark=at position 0.57 with {\arrow{Triangle[scale=0.65, reversed]}}
    },
    thick
  },
fermionloop/.style={
    postaction={decorate},
    decoration={
      markings},
    thick},
momentum/.style={
    postaction={decorate},
    decoration={
      markings,
      mark=at position 0.28 with {\arrow{Triangle[scale=0.7]}},
      mark=at position 0.78 with {\arrow{Triangle[scale=0.7]}},
    }
  },
 photon/.style={
  thick,
  decorate,                  
  decoration={
    snake,                   
    amplitude=1.5pt,           
    segment length=5pt       
  }}]

\coordinate (A) at (0.5,0);
\coordinate (B) at (1.5,0);
\coordinate (X) at (0,0); 
\coordinate (X1) at (2,0);
\filldraw[white] (X) circle[radius=0.5cm]; 
\filldraw[white] (X1) circle[radius=0.5cm]; 
\draw[fermionloop,line width=1.2pt] (X)  circle[radius=0.5cm];
\draw[fermionloop,line width=1.2pt] (X1)  circle[radius=0.5cm];
\path[momentum] (X)  circle[radius=0.5cm];
\path[momentum] (X1)  circle[radius=0.5cm];
\filldraw[fill=black] (-0.5,0) node[left=1.5pt] {$\cdots$} circle[radius=1.2pt];
\filldraw[fill=black] (0.5,0) node[right=1.5pt] {} circle[radius=1.2pt];
\filldraw[fill=black] (1.5,0)node[left=1.5pt] {} circle[radius=1.2pt];
\filldraw[fill=black] (2.5,0) node[right=1.5pt] {$\cdots$} circle[radius=1.2pt];
\draw[photon,line width=1.2pt] (A) --(B);
\draw[line width=1.2pt] (-0.5,0) circle[radius=0.12];
\draw[line width=1.2pt] (2.5,0) circle[radius=0.12];
\end{tikzpicture}}}=
0.
\end{equation} Indeed, integrating over the two non-circled vertices in the middle
yields a non-zero net energy flowing out, so this diagram vanishes.
Therefore, like the case in scalar field, we are instructed to retain
only those diagrams in which one side of the cut contains exclusively
non-circled vertices, while the other side contains only circled ones,
with the external energy flowing from the non-circled side to the
circled side, just like the second diagram on the second line of Eq.~(\ref{eq:cutting_diagram}).

Applying these rules to the infinite series in Eq.~(\ref{eq:RPA_series})
and performing some algebra, again we obtain the optical theorem for
the DM self-energy, \begin{align}
& 2~\mathrm{Im}\left(
\begin{tikzpicture}[baseline={(0,0.18)}, scale=1,       
  thick,
  DMfermion/.style={
    draw=gray!70,
    postaction={decorate},
    decoration={
      markings,
      mark=at position 0.57 with {\arrow{Triangle[scale=0.65, reversed]}}
    },
    thick
  },
  photon/.style={
    dashed,
    thick
  }]
  \coordinate (A) at (-1.5,0);
  \coordinate (B) at (-0.8,0);
  \coordinate (C) at (0.8,0);
  \coordinate (D) at (1.5,0);
  \coordinate (X) at (0,0.7); 
  \draw[DMfermion,line width=1.2pt] (A) -- (B) node[midway,above=2pt,black] {\small$\mathbf{p}_{\chi}$};
  \draw[DMfermion,line width=1.2pt] (B) -- (C);
  \draw[DMfermion,line width=1.2pt] (C) -- (D) node[midway,above=2pt,black] {\small$\mathbf{p}_{\chi}$};
  \filldraw[fill=black] (B) circle[radius=1.2pt];
  \filldraw[fill=black] (C) circle[radius=1.2pt];
  \draw[photon,line width=1.2pt] (B) to[out=90,in=180] coordinate[pos=0.7] (X1) (X);
  \draw[photon,line width=1.2pt] (X) to[out=0,in=90] coordinate[pos=0.7] (X2) (C);
\end{tikzpicture}
+
\begin{tikzpicture}[baseline={(0,0.18)}, scale=1,
  thick,
  DMfermion/.style={
    draw=gray!70,
    postaction={decorate},
    decoration={
      markings,
      mark=at position 0.57 with {\arrow{Triangle[scale=0.65, reversed]}}
    },
    thick
  },
  photon/.style={
    dashed,
    thick
  },
  fermionloop/.style={
    postaction={decorate},
    decoration={
      markings,
      mark=at position 0.79 with {\arrow{Triangle[scale=0.65]}}
    },
    thick
  },
  momentum/.style={
    postaction={decorate},
    decoration={
      markings,
      mark=at position 0.28 with {\arrow{Triangle[scale=0.65]}}
    }
  }]
  \coordinate (A) at (-1.5,0);
  \coordinate (B) at (-0.8,0);
  \coordinate (C) at (0.8,0);
  \coordinate (D) at (1.5,0);
  \coordinate (X) at (0,0.7); 
  \draw[DMfermion,line width=1.2pt] (A) -- (B) node[midway,above=2pt,black] {\small$\mathbf{p}_{\chi}$};
  \draw[DMfermion,line width=1.2pt] (B) -- (C);
  \draw[DMfermion,line width=1.2pt] (C) -- (D) node[midway,above=2pt,black] {\small$\mathbf{p}_{\chi}$};
  \filldraw[fill=black] (B) circle[radius=1.2pt];
  \filldraw[fill=black] (C) circle[radius=1.2pt];
  \draw[photon,line width=1.2pt] (B) to[out=90,in=180] coordinate[pos=0.7] (X1) (X);
  \draw[photon,line width=1.2pt] (X) to[out=0,in=90] coordinate[pos=0.7] (X2) (C);
  \filldraw[white] (X) circle[radius=0.3cm]; 
  \draw[fermionloop,line width=1.2pt] (X) circle[radius=0.3cm];
  \path[momentum] (X) circle[radius=0.3cm];
  \filldraw[fill=black] (-0.29,0.68) circle[radius=1.2pt];
  \filldraw[fill=black] (0.29,0.68) circle[radius=1.2pt];
\end{tikzpicture}
+
\begin{tikzpicture}[baseline={(0,0.18)}, scale=1,
  thick,
  DMfermion/.style={
    draw=gray!70,
    postaction={decorate},
    decoration={
      markings,
      mark=at position 0.57 with {\arrow{Triangle[scale=0.65, reversed]}}
    },
    thick
  },
  photon/.style={
    dashed,
    thick
  },
  photon1/.style={
    thick,
    decorate,
    decoration={
      snake,
      amplitude=1.5pt,
      segment length=4pt
    }
  },
  fermionloop/.style={
    postaction={decorate},
    decoration={
      markings,
      mark=at position 0.79 with {\arrow{Triangle[scale=0.65]}}
    },
    thick
  },
  momentum/.style={
    postaction={decorate},
    decoration={
      markings,
      mark=at position 0.28 with {\arrow{Triangle[scale=0.65]}}
    }
  }]
  \coordinate (A) at (-1.8,0);
  \coordinate (B) at (-1,0);
  \coordinate (C) at (1,0);
  \coordinate (D) at (1.8,0);
  \coordinate (X) at (-0.6,0.7);
  \coordinate (X1) at (0.6,0.7);
  \draw[DMfermion,line width=1.2pt] (A) -- (B) node[midway,above=2pt,black] {\small$\mathbf{p}_{\chi}$};
  \draw[DMfermion,line width=1.2pt] (B) -- (C);
  \draw[DMfermion,line width=1.2pt] (C) -- (D) node[midway,above=2pt,black] {\small$\mathbf{p}_{\chi}$};
  \filldraw[fill=black] (B) circle[radius=1.2pt];
  \filldraw[fill=black] (C) circle[radius=1.2pt];
  \draw[photon,line width=1.2pt] (B) to[out=90,in=180] (X);
  \draw[photon1,line width=1.2pt] (X) -- (X1);
  \draw[photon,line width=1.2pt] (C) to[out=90,in=0] (X1);
  \filldraw[white] (X) circle[radius=0.3cm];
  \filldraw[white] (X1) circle[radius=0.3cm];
  \draw[fermionloop,line width=1.2pt] (X) circle[radius=0.3cm];
  \draw[fermionloop,line width=1.2pt] (X1) circle[radius=0.3cm];
  \path[momentum] (X) circle[radius=0.3cm];
  \path[momentum] (X1) circle[radius=0.3cm];
  \filldraw[fill=black] (-0.86,0.55) circle[radius=1.2pt];
  \filldraw[fill=black] (0.86,0.55) circle[radius=1.2pt];
\end{tikzpicture}
+\cdots \right)\nonumber \\[10pt]
={} & 
\begin{tikzpicture}[baseline={(0,0.18)}, scale=1,
  thick,
  DMfermion/.style={
    draw=gray!70,
    postaction={decorate},
    decoration={
      markings,
      mark=at position 0.57 with {\arrow{Triangle[scale=0.65, reversed]}}
    },
    thick
  },
  photon/.style={
    dashed,
    thick
  },
  fermionloop/.style={
    postaction={decorate},
    decoration={
      markings,
      mark=at position 0.82 with {\arrow{Triangle[scale=0.65]}}
    },
    thick
  },
  momentum/.style={
    postaction={decorate},
    decoration={
      markings,
      mark=at position 0.32 with {\arrow{Triangle[scale=0.65]}}
    }
  }]
  \coordinate (A) at (-1.5,0);
  \coordinate (B) at (-0.8,0);
  \coordinate (C) at (0.8,0);
  \coordinate (D) at (1.5,0);
  \coordinate (X) at (0,0.7); 
  \draw[DMfermion,line width=1.2pt] (A) -- (B) node[midway,above=2pt,black] {\small$\mathbf{p}_{\chi}$};
  \draw[DMfermion,line width=1.2pt] (B) -- (C);
  \draw[DMfermion,line width=1.2pt] (C) -- (D) node[midway,above=2pt,black] {\small$\mathbf{p}_{\chi}$};
  \filldraw[fill=black] (B) circle[radius=1.2pt];
  \filldraw[fill=black] (C) circle[radius=1.2pt];
  \draw[photon,line width=1.2pt] (B) to[out=90,in=180] coordinate[pos=0.7] (X1) (X);
  \draw[photon,line width=1.2pt] (X) to[out=0,in=90] coordinate[pos=0.7] (X2) (C);
  \filldraw[white] (X) circle[radius=0.3cm]; 
  \draw[fermionloop,line width=1.2pt] (X) circle[radius=0.3cm];
  \path[momentum] (X) circle[radius=0.3cm];
  \filldraw[fill=black] (-0.29,0.68) circle[radius=1.2pt];
  \filldraw[fill=black] (0.29,0.68) circle[radius=1.2pt];
  \draw[gray!90,dotted,line width=1.5pt] ($(X)+(0,-1.1)$) -- ($(X)+(0,0.8)$);
\end{tikzpicture}
+
\begin{tikzpicture}[baseline={(0,0.18)}, scale=1,
  thick,
  DMfermion/.style={
    draw=gray!70,
    postaction={decorate},
    decoration={
      markings,
      mark=at position 0.57 with {\arrow{Triangle[scale=0.65, reversed]}}
    },
    thick
  },
  photon/.style={
    dashed,
    thick
  },
  photon1/.style={
    thick,
    decorate,
    decoration={
      snake,
      amplitude=1.5pt,
      segment length=4pt
    }
  },
  fermionloop/.style={
    postaction={decorate},
    decoration={
      markings,
      mark=at position 0.82 with {\arrow{Triangle[scale=0.65]}}
    },
    thick
  },
  momentum/.style={
    postaction={decorate},
    decoration={
      markings,
      mark=at position 0.32 with {\arrow{Triangle[scale=0.65]}}
    }
  }]
  \coordinate (A) at (-1.8,0);
  \coordinate (B) at (-1,0);
  \coordinate (C) at (1,0);
  \coordinate (D) at (1.8,0);
  \coordinate (X) at (-0.6,0.7);
  \coordinate (X1) at (0.6,0.7);
  \draw[DMfermion,line width=1.2pt] (A) -- (B) node[midway,above=2pt,black] {\small$\mathbf{p}_{\chi}$};
  \draw[DMfermion,line width=1.2pt] (B) -- (C);
  \draw[DMfermion,line width=1.2pt] (C) -- (D) node[midway,above=2pt,black] {\small$\mathbf{p}_{\chi}$};
  \filldraw[fill=black] (B) circle[radius=1.2pt];
  \filldraw[fill=black] (C) circle[radius=1.2pt];
  \draw[photon,line width=1.2pt] (B) to[out=90,in=180] (X);
  \draw[photon1,line width=1.2pt] (X) -- (X1);
  \draw[photon,line width=1.2pt] (C) to[out=90,in=0] (X1);
  \filldraw[white] (X) circle[radius=0.3cm];
  \filldraw[white] (X1) circle[radius=0.3cm];
  \draw[fermionloop,line width=1.2pt] (X) circle[radius=0.3cm];
  \draw[fermionloop,line width=1.2pt] (X1) circle[radius=0.3cm];
  \path[momentum] (X) circle[radius=0.3cm];
  \path[momentum] (X1) circle[radius=0.3cm];
  \filldraw[fill=black] (-0.86,0.55) circle[radius=1.2pt];
  \filldraw[fill=black] (0.86,0.55) circle[radius=1.2pt];
  \draw[gray!90,dotted,line width=1.5pt] ($(X)+(0,-1.1)$) -- ($(X)+(0,0.8)$);
\end{tikzpicture}
+ 
\begin{tikzpicture}[baseline={(0,0.18)}, scale=1,
  thick,
  DMfermion/.style={
    draw=gray!70,
    postaction={decorate},
    decoration={
      markings,
      mark=at position 0.57 with {\arrow{Triangle[scale=0.65, reversed]}}
    },
    thick
  },
  photon/.style={
    dashed,
    thick
  },
  photon1/.style={
    thick,
    decorate,
    decoration={
      snake,
      amplitude=1.5pt,
      segment length=4pt
    }
  },
  fermionloop/.style={
    postaction={decorate},
    decoration={
      markings,
      mark=at position 0.82 with {\arrow{Triangle[scale=0.65]}}
    },
    thick
  },
  momentum/.style={
    postaction={decorate},
    decoration={
      markings,
      mark=at position 0.32 with {\arrow{Triangle[scale=0.65]}}
    }
  }]
  \coordinate (A) at (-1.8,0);
  \coordinate (B) at (-1,0);
  \coordinate (C) at (1,0);
  \coordinate (D) at (1.8,0);
  \coordinate (X) at (-0.6,0.7);
  \coordinate (X1) at (0.6,0.7);
  \draw[DMfermion,line width=1.2pt] (A) -- (B) node[midway,above=2pt,black] {\small$\mathbf{p}_{\chi}$};
  \draw[DMfermion,line width=1.2pt] (B) -- (C);
  \draw[DMfermion,line width=1.2pt] (C) -- (D) node[midway,above=2pt,black] {\small$\mathbf{p}_{\chi}$};
  \filldraw[fill=black] (B) circle[radius=1.2pt];
  \filldraw[fill=black] (C) circle[radius=1.2pt];
  \draw[photon,line width=1.2pt] (B) to[out=90,in=180] (X);
  \draw[photon1,line width=1.2pt] (X) -- (X1);
  \draw[photon,line width=1.2pt] (C) to[out=90,in=0] (X1);
  \filldraw[white] (X) circle[radius=0.3cm];
  \filldraw[white] (X1) circle[radius=0.3cm];
  \draw[fermionloop,line width=1.2pt] (X) circle[radius=0.3cm];
  \draw[fermionloop,line width=1.2pt] (X1) circle[radius=0.3cm];
  \path[momentum] (X) circle[radius=0.3cm];
  \path[momentum] (X1) circle[radius=0.3cm];
  \filldraw[fill=black] (-0.86,0.55) circle[radius=1.2pt];
  \filldraw[fill=black] (0.86,0.55) circle[radius=1.2pt];
  \draw[gray!90,dotted,line width=1.5pt] ($(X1)+(0,-1.1)$) -- ($(X1)+(0,0.8)$);
\end{tikzpicture}
+\cdots \nonumber \\[10pt]
={} & 
\sum_{\mathbf{p}'_{\chi}}\sum_{i>F}\sum_{j\leq F}~\left|
\vcenter{\hbox{\tikzset{every picture/.style={line width=0.75pt}}
\begin{tikzpicture}[
  thick,
  DMfermion/.style={draw=gray!70,
    postaction={decorate},
    decoration={
      markings,
      mark=at position 0.57 with {\arrow{Triangle[scale=0.65,reversed]}}
    },
    thick
  },
   fermion/.style={
    postaction={decorate},
    decoration={
      markings,
      mark=at position 0.57 with {\arrow{Triangle[scale=0.65]}}
    },
    thick
  },
  photon/.style={
    dashed,          
    thick
  }]
\coordinate (A) at (-1.1,-0.75);
\coordinate (B) at (-0.65,0);
\coordinate (C) at (-1.1,0.75);
\coordinate (D) at (0.65,0);
\coordinate (E) at (1.1,0.75);
\coordinate (F) at (1.1,-0.75);
\coordinate (H) at (0.85,0);
\coordinate (I) at (1.3,-0.75);
\draw[DMfermion,line width=1.2pt] (B) node[anchor=south east, xshift=-13pt, yshift=10pt] {$\mathbf{p'}_{\chi}$} -- (A) node[midway,below] {};
\draw[DMfermion,line width=1.2pt] (C) node[above] {$$} -- (B) node[anchor=south east, xshift=-15pt, yshift=-23pt] {$\mathbf{p}_{\chi}$};
\draw[
    double,          
    double distance=0.5pt,  
    dashed,          
    line width=0.8pt 
  ] (B) -- (D) node[midway, above] {};
\draw[fermion,line width=1.2pt] (D)node[anchor=north west, xshift=16pt, yshift=30pt] {$i$} -- (E) node[midway,below] {$$} node[right] {};
\draw[fermion,line width=1.2pt] (F) -- (D) node[anchor=north west, xshift=16pt, yshift=-12pt] {$j$};
\draw[line width=1.2pt,
    dashed, color=gray,
    decoration={markings, mark=at position 0.57 with {\arrow{Triangle[scale=0.65]}}},
    postaction={decorate}
  ] 
  (H) -- (I);
\filldraw[fill=black] (B) circle[radius=1.2pt];
\filldraw[fill=black] (D) circle[radius=1.2pt];
\end{tikzpicture}}}\right|^{2},
\label{eq:cutting_diagram_RPA}
\end{align}where the double line represents the screened DM-electron interaction,
\begin{align}
\begin{tikzpicture}[baseline={(0,-0.1)}, scale=1]
\coordinate (B) at (-0.6,0);
\coordinate (D) at (0.6,0);
\draw[
    double,          
    double distance=0.7pt,  
    dashed,          
    line width=0.8pt 
  ] (B) -- (D) node[midway, above] {};
\end{tikzpicture}&~=~
\begin{tikzpicture}[baseline={(0,-0.1)}, scale=1,
  thick,
  photon/.style={
    dashed,
    thick
  }]
\coordinate (B) at (-0.6,0);
\coordinate (D) at (0.6,0);
\draw[photon,line width=1.2pt] (B)--(D) node[midway, above] {};
\end{tikzpicture}~+~
\begin{tikzpicture}[baseline={(0,-0.1)}, scale=1,
  thick,
  DMfermion/.style={
    draw=gray!70,
    postaction={decorate},
    decoration={
      markings,
      mark=at position 0.57 with {\arrow{Triangle[scale=0.65, reversed]}}
    },
    thick
  },
  photon/.style={
    dashed,
    thick
  },
  photon1/.style={
    thick,
    decorate,
    decoration={
      snake,
      amplitude=1.5pt,
      segment length=4pt
    }
  },
  fermionloop/.style={
    postaction={decorate},
    decoration={
      markings,
      mark=at position 0.79 with {\arrow{Triangle[scale=0.65]}}
    },
    thick
  },
  momentum/.style={
    postaction={decorate},
    decoration={
      markings,
      mark=at position 0.28 with {\arrow{Triangle[scale=0.65]}}
    }
  }]
  \coordinate (A) at (-1,0);
  \coordinate (B) at (-0.4,0);
  \coordinate (C) at (0.4,0);
  \coordinate (D) at (1,0);
  \coordinate (X) at (0,0);
  \coordinate (X1) at (0.6,0.7);
  \draw[photon,line width=1.2pt] (A) -- (B) node[midway,above=2pt,black] {};
  \draw[photon1,line width=1.2pt] (C) -- (D);
  \filldraw[white] (X) circle[radius=0.4cm];
  \draw[fermionloop,line width=1.2pt] (X) circle[radius=0.4cm];
  \path[momentum] (X) circle[radius=0.4cm];
 \filldraw[fill=black] (B) circle[radius=1.2pt];
 \filldraw[fill=black] (C) circle[radius=1.2pt];
\end{tikzpicture}~+~
\begin{tikzpicture}[baseline={(0,-0.1)}, scale=1,
  thick,
  DMfermion/.style={
    draw=gray!70,
    postaction={decorate},
    decoration={
      markings,
      mark=at position 0.57 with {\arrow{Triangle[scale=0.65, reversed]}}
    },
    thick
  },
  photon/.style={
    dashed,
    thick
  },
  photon1/.style={
    thick,
    decorate,
    decoration={
      snake,
      amplitude=1.5pt,
      segment length=4pt
    }
  },
  fermionloop/.style={
    postaction={decorate},
    decoration={
      markings,
      mark=at position 0.79 with {\arrow{Triangle[scale=0.65]}}
    },
    thick
  },
  momentum/.style={
    postaction={decorate},
    decoration={
      markings,
      mark=at position 0.28 with {\arrow{Triangle[scale=0.65]}}
    }
  }]
  \coordinate (A) at (-1,0);
  \coordinate (B) at (-0.4,0);
  \coordinate (C) at (0.4,0);
  \coordinate (D) at (1,0);
  \coordinate (E) at (1.8,0);
  \coordinate (F) at (2.4,0);
  \coordinate (X) at (0,0);
  \coordinate (X1) at (1.4,0);
  \draw[photon,line width=1.2pt] (A) -- (B) node[midway,above=2pt,black] {};
  \draw[photon1,line width=1.2pt] (C) -- (D);
  \draw[photon1,line width=1.2pt] (E) -- (F);
  \filldraw[white] (X) circle[radius=0.4cm];
  \filldraw[white] (X1) circle[radius=0.4cm];
  \draw[fermionloop,line width=1.2pt] (X) circle[radius=0.4cm];
  \draw[fermionloop,line width=1.2pt] (X1) circle[radius=0.4cm];
  \path[momentum] (X) circle[radius=0.4cm];
  \path[momentum] (X1) circle[radius=0.4cm];
 \filldraw[fill=black] (B) circle[radius=1.2pt];
 \filldraw[fill=black] (C) circle[radius=1.2pt];
 \filldraw[fill=black] (D) circle[radius=1.2pt];
 \filldraw[fill=black] (E) circle[radius=1.2pt];
 \end{tikzpicture}~+~\cdots \nonumber \\
&~=~ 
\frac{
\begin{tikzpicture}[baseline={(0,-0.2)}, scale=1,
photon/.style={
    dashed,
    thick
  }]
\coordinate (B) at (-0.6,0);
\coordinate (D) at (0.6,0);
\draw[photon,line width=1.2pt] (B) -- (D) node[midway, above] {};
\end{tikzpicture}}{1-\begin{tikzpicture}[baseline={(0,-0.1)}, scale=1,
  thick,
  DMfermion/.style={
    draw=gray!70,
    postaction={decorate},
    decoration={
      markings,
      mark=at position 0.57 with {\arrow{Triangle[scale=0.65, reversed]}}
    },
    thick
  },
  photon/.style={
    dashed,
    thick
  },
  photon1/.style={
    thick,
    decorate,
    decoration={
      snake,
      amplitude=1.5pt,
      segment length=4pt
    }
  },
  fermionloop/.style={
    postaction={decorate},
    decoration={
      markings,
      mark=at position 0.79 with {\arrow{Triangle[scale=0.65]}}
    },
    thick
  },
  momentum/.style={
    postaction={decorate},
    decoration={
      markings,
      mark=at position 0.28 with {\arrow{Triangle[scale=0.65]}}
    }
  }]
  \coordinate (A) at (-1,0);
  \coordinate (B) at (-0.3,0);
  \coordinate (C) at (0.3,0);
  \coordinate (D) at (1,0);
  \coordinate (X) at (0,0);
  \draw[photon1,line width=1.2pt] (C) -- (D);
  \filldraw[white] (X) circle[radius=0.3cm];
  \draw[fermionloop,line width=1.2pt] (X) circle[radius=0.3cm];
  \path[momentum] (X) circle[radius=0.3cm];
\filldraw[fill=black] (B) circle[radius=1.2pt];
\filldraw[fill=black] (C) circle[radius=1.2pt];
\end{tikzpicture}}\nonumber \\[10pt]
&~=~
\frac{
\begin{tikzpicture}[baseline={(0,-0.2)}, scale=1,
photon/.style={
    dashed,
    thick
  }]
\coordinate (B) at (-0.6,0);
\coordinate (D) at (0.6,0);
\draw[photon,line width=1.2pt] (B) -- (D) node[midway, above] {};
\end{tikzpicture}}{\epsilon_{\mathrm{RPA}}},
\end{align}with 
\begin{eqnarray}
\epsilon_{\mathrm{RPA}}\left(\mathbf{Q},\,\omega\right) & = & 1-\frac{1}{V}\frac{4\pi\alpha}{\left|\mathbf{Q}\right|^{2}}\sum_{i,j}\frac{\left|\int\mathrm{d}^{3}x\,u_{i}^{*}\left(\mathbf{x}\right)e^{i\mathbf{Q}\cdot\mathbf{x}}u_{j}\left(\mathbf{x}\right)\right|^{2}}{\varepsilon_{i}-\varepsilon_{j}-\omega-i0^{+}}\left(f_{i}-f_{j}\right)\label{eq:dielectric function}
\end{eqnarray}
being the relevant dielectric function for the four-momentum $\left(\omega,\,\mathbf{Q}\right)$
transferred from DM to the medium ($f_{k}=0,\,1$ represents the occupation
number of $k$-th KS orbital, if the spin degrees of freedom are ignored).
Then the decay rate of DM in medium reads as 
\begin{eqnarray}
\Gamma_{p_{\chi}} & = & \sum_{\mathbf{Q}}\sum_{i>F}\sum_{j\leq F}\frac{2\pi}{V^{2}}\delta\left(\varepsilon_{\mathbf{p}_{\chi}}-\varepsilon_{\mathbf{p}'_{\chi}}-\varepsilon_{i}+\varepsilon_{j}\right)\left|\frac{-ig_{\chi}g_{e}\int\mathrm{d}^{3}x\,u_{i}^{*}\left(\mathbf{x}\right)e^{i\mathbf{\mathbf{Q}}\cdot\mathbf{x}}u_{j}\left(\mathbf{x}\right)}{\left(\left|\mathbf{Q}\right|^{2}+m_{A'}^{2}\right)\epsilon_{\mathrm{RPA}}\left(\mathbf{Q},\varepsilon_{\mathbf{p}_{\chi}}-\varepsilon_{\mathbf{p}'_{\chi}}\right)}\right|^{2},\nonumber \\
\label{eq:decay rate}
\end{eqnarray}
where the sum is over the momentum transferred to the medium $\mathbf{\mathbf{Q}}$,
$\varepsilon_{\mathbf{p}'_{\chi}}=\left(\mathbf{p}_{\chi}-\mathbf{Q}\right)^{2}/\left(2m_{\chi}\right)$
represents the energy of the decayed DM particle. If the two spin
orientations are taken into account, a factor of $2$ should be added
in both $f_{k}$ in Eq.~(\ref{eq:dielectric function}) and in the
decay rate of  Eq.~(\ref{eq:decay rate}). Although this expression
can also be derived using Fermi’s golden rule, the screening effect
$\epsilon_{\mathrm{RPA}}\left(\mathbf{\mathbf{Q}},\omega\right)$
must be introduced phenomenologically in such an approach. 

On the other hand, from the perspective of the linear response theory,
the dielectric function $\epsilon\left(\mathbf{\mathbf{Q}},\omega\right)$
encodes the linear response of the system to a longitudinal perturbation,
whether induced by a DM particle or an external electromagnetic field.
In this context, a compact formalism for the DM scattering rate in
solids has been derived in Refs.~\citep{Hochberg:2021pkt,Knapen:2021run}:
\begin{eqnarray}
\Gamma\left(\mathbf{p}_{\chi}\right) & = & \frac{2}{V}\sum_{\mathbf{Q}}\frac{\left|\mathbf{Q}\right|^{2}}{4\pi\alpha}\left|\frac{-ig_{\chi}g_{e}}{\left(\left|\mathbf{Q}\right|^{2}+m_{A'}^{2}\right)}\right|^{2}\mathrm{Im}\left[\frac{-1}{\epsilon\left(\mathbf{\mathbf{Q}},\varepsilon_{\mathbf{p}_{\chi}}-\varepsilon_{\mathbf{p}'_{\chi}}\right)}\right],\label{eq:ELF_scattering}
\end{eqnarray}
where the dielectric function $\epsilon\left(\mathbf{\mathbf{Q}},\omega\right)$
and hence the\emph{ energy loss function} (ELF) $\mathrm{Im}\left[-\epsilon^{-1}\left(\mathbf{\mathbf{Q}},\omega\right)\right]$
can be obtained from first-principles calculations, phenomenological
fittings, or experimental measurements. For example, the well-known
Lindhard dielectric function arises from applying the RPA to the simplest
model of a free electron gas. Notably, when the RPA is specifically
applied to the KS orbitals, the resulting DM scattering rate coincides
with the decay rate given in Eq.~(\ref{eq:decay rate}). And more
importantly, by directly calibrating the spectrum using electromagnetic
probes (\emph{e.g.}, infrared spectroscopy, X-ray scattering, and
\emph{electron energy-loss spectroscopy} (EELS)) in the kinematically
accessible regime relevant for the DM scattering, one can circumvent
the need for electronic wavefunctions to compute the dielectric function
$\epsilon\left(\mathbf{\mathbf{Q}},\omega\right)$, thereby removing
theoretical uncertainties associated with the material physics of
the target system~\citep{Hochberg:2021pkt,Knapen:2021run}.

\section{\label{sec:Conclusions}Summary and discussion}

Due to the DM-electron interaction, DM particles in solids acquire
a finite lifetime and can decay into lower-energy DM particles and
electron-hole pairs. To compute the decay rate, one can utilize the
self-energy, defined as the sum of 1PI insertions into the propagator.
The imaginary parts of the relevant loop diagrams determine the decay
rate. While Cutkosky’s cutting rules provide an efficient method for
simplifying these calculations in relativistic QED, analogous rules
for Coulomb interactions beyond the homogeneous electron gas approximation
in solids remain underdeveloped. In this paper, we fill this gap by
providing a proof for the cutting rules for Coulomb and instantaneous
DM-electron interactions. In addition, the relevant electron wavefunctions
are obtained using the DFT approach, rather than naive homogeneous
electron plane-wave wavefunctions.

While the derived cutting rules are technically practical, they have
some notable differences compared to those for relativistic scalar
theories and QED, where the intermediate particle propagators are
constructed at physical masses (or at the pole locations), so as to
make the perturbation theory more plausible. In contrast, the cutting
rules derived in this work use the KS orbitals as the starting point
for the perturbation approach. Although the Kohn-Sham DFT method provides
reliable predictions for crystal structures, lattice constants, and
atomic geometries, which are essential for computing electronic properties,
the eigenvalues from KS equations do not directly correspond to quasi-particle
excitation energies and thus results in a underestimation of bandgaps
in semiconductors and insulators. In this sense, the electron-hole
propagator based on the KS equation merely serves as a reasonable
starting point for the one-shot perturbation approach. In order to
obtain a more accurate description of quasi-particles energies and
electronic excitations, one can resort to self-consistent methods
such as\emph{ $GW$} approximation~\citep{Hedin:1965zza}. However,
the present framework still achieves a good balance between accuracy
and efficiency.

In certain cases, it is essential to resum specific classes of Feynman
diagrams to all orders in order to obtain an accurate description
of phenomena in condensed matter physics. For example, employing the
unscreened Coulomb interaction in the HF self-energy completely neglects
correlation effects and treats exchange at a mean-field level using
the bare Coulomb interaction. This can lead to an inaccurate description
of key quantities such as effective masses and quasi-particle lifetimes.
Incorporating screening effects addresses the inherent limitations
of HF theory that arise from the long-range nature of the Coulomb
interaction, which involves insertion of an infinite number of polarized
electron bubbles into the Coulomb propagator. The cutting rules derived
in this work can be implemented even when the calculation of electron
and DM particle self-energies involves a resummation of Feynman diagrams
to all orders, rather than being restricted to a loop-by-loop perturbative
expansion. These cutting rules therefore provide a powerful and systematic
framework for analyzing interacting many-body systems, including phenomena
such as dynamical screening and plasmon excitations---effects that
cannot be captured by standard two-body scattering processes or within
a picture of non-interacting single-particle states.

\appendix

\renewcommand{\theequation}{A.\arabic{equation}}
\begin{acknowledgments}
The work of Z.L.L is supported by the National Natural Science Foundation
of China under No.~12575117, and the work of F.Z is supported by
the National Natural Science Foundation of China under No.~12374054.. 
\end{acknowledgments}

\section{\label{sec:appendix1}Feynman rules for DM-electron interaction in
solids}

In this appendix we provide a collection of Feynman rules associated
with non-relativistic DM particles and KS electrons and holes in solids:
\begin{subequations}
\begin{align}
\text{outgoing DM leg:} \quad 
\vcenter{\hbox{\tikzset{every picture/.style={line width=0.75pt}}
\begin{tikzpicture}[x=0.55pt,y=0.55pt,yscale=-1]
\begin{scope}[shift={(150,100)}]
  \draw[gray,line width=1.2pt,
    decoration={markings, mark=at position 0.57 with {\arrow{Triangle[scale=0.65]}}},
    postaction={decorate}
  ] 
  (100,0) -- (0,0) 
    node[midway,above,black] {$\mathbf{p}'_{\chi}$};
  \filldraw[black] (100,0) circle[radius=1.2pt]
    node[below=2pt] {$z$};
\end{scope}
\end{tikzpicture}}} 
&= \frac{e^{-i\mathbf{p}'_{\chi}\cdot\mathbf{z}}}{\sqrt{V}}e^{i\varepsilon_{\mathbf{p}'_{\chi}}t_{z}} 
\label{eq:outgoing_DM} \\
\text{incoming DM leg:} \quad 
\vcenter{\hbox{\tikzset{every picture/.style={line width=0.75pt}}
\begin{tikzpicture}[x=0.55pt,y=0.55pt,yscale=-1]
\begin{scope}[shift={(150,100)}]
  \draw[gray,line width=1.2pt,
    decoration={markings, mark=at position 0.57 with {\arrow{Triangle[scale=0.65]}}},
    postaction={decorate}
  ] 
  (0,0) -- (100,0) 
    node[midway,above,black] {$\mathbf{p}_{\chi}$};
  \filldraw[black] (100,0) circle[radius=1.2pt]
    node[below=2pt] {$z$};
\end{scope}
\end{tikzpicture}}} 
&= \frac{e^{i\mathbf{p}_{\chi}\cdot\mathbf{z}}}{\sqrt{V}}e^{-i\varepsilon_{\mathbf{p}_{\chi}}t_{z}}
\label{eq:incoming_DM} \\
\text{outgoing electron leg:} \quad 
\vcenter{\hbox{\tikzset{every picture/.style={line width=0.75pt}}
\begin{tikzpicture}[x=0.55pt,y=0.55pt,yscale=-1]
\begin{scope}[shift={(150,100)}]
  \draw[line width=1.2pt,
    decoration={markings, mark=at position 0.57 with {\arrow{Triangle[scale=0.65]}}},
    postaction={decorate}
  ] 
  (100,0) -- (0,0) 
    node[midway,above] {$i$};
  \filldraw[black] (100,0) circle[radius=1.2pt]
    node[below=2pt] {$z$};
\end{scope}
\end{tikzpicture}}} 
&= u_{i}^{*}\left(\mathbf{z}\right)e^{i\varepsilon_{i}t_{z}} 
\label{eq:outgoing_e} \\
\text{incoming electron leg:} \quad 
\vcenter{\hbox{\tikzset{every picture/.style={line width=0.75pt}}
\begin{tikzpicture}[x=0.55pt,y=0.55pt,yscale=-1]
\begin{scope}[shift={(150,100)}]
  \draw[line width=1.2pt,
    decoration={markings, mark=at position 0.57 with {\arrow{Triangle[scale=0.65]}}},
    postaction={decorate}
  ] 
  (0,0) -- (100,0) 
    node[midway,above] {$j$};
  \filldraw[black] (100,0) circle[radius=1.2pt]
    node[below=2pt] {$z$};
\end{scope}
\end{tikzpicture}}} 
&= u_{j}\left(\mathbf{z}\right)e^{-i\varepsilon_{j}t_{z}} 
\label{eq:incoming_e} \\
\text{outgoing hole leg:} \quad 
\vcenter{\hbox{\tikzset{every picture/.style={line width=0.75pt}}
\begin{tikzpicture}[x=0.55pt,y=0.55pt,yscale=-1]
\begin{scope}[shift={(150,100)}]
  \draw[line width=1.2pt,
    decoration={markings, mark=at position 0.57 with {\arrow{Triangle[scale=0.65]}}},
    postaction={decorate}
  ] 
  (0,0) -- (100,0)
    node[midway, above] {$m$};
  \draw[line width=1.2pt,
    dashed, color=gray,
    decoration={markings, mark=at position 0.57 with {\arrow{Triangle[scale=0.65, reversed]}}},
    postaction={decorate}
  ] 
  (15,10) -- (85,10);
  \filldraw[black] (100,0) circle[radius=1.2pt]
    node[below=2pt] {$z$};
\end{scope}
\end{tikzpicture}}} 
&= \left(-1\right)u_{m}\left(\mathbf{z}\right)e^{-i\varepsilon_{m}t_{z}} 
\label{eq:outgoing_h} \\
\text{incoming hole leg:} \quad 
\vcenter{\hbox{\tikzset{every picture/.style={line width=0.75pt}}
\begin{tikzpicture}[x=0.55pt,y=0.55pt,yscale=-1]
\begin{scope}[shift={(150,100)}]
  \draw[line width=1.2pt,
    decoration={markings, mark=at position 0.57 with {\arrow{Triangle[scale=0.65, reversed]}}},
    postaction={decorate}
  ] 
  (100,0) -- (0,0)
    node[midway, above] {$n$};
  \draw[line width=1.2pt,
    dashed, color=gray,
    decoration={markings, mark=at position 0.57 with {\arrow{Triangle[scale=0.65]}}},
    postaction={decorate}
  ] 
  (85,10) -- (15,10);
  \filldraw[black] (100,0) circle[radius=1.2pt]
    node[below=2pt] {$z$};
\end{scope}
\end{tikzpicture}}} 
&= \left(-1\right)u_{n}^{*}\left(\mathbf{z}\right)e^{i\varepsilon_{n}t_{z}} 
\label{eq:incoming_h} \\
\text{electron internal line:} \quad 
\vcenter{\hbox{\tikzset{every picture/.style={line width=0.75pt}}
\begin{tikzpicture}[x=0.55pt,y=0.55pt,yscale=-1,baseline={([yshift=20pt]current bounding box.center)}]
\begin{scope}[shift={(150,100)}]
\draw[line width=1.2pt,
    decoration={markings, mark=at position 0.57 with {\arrow{Triangle[scale=0.65]}}},
    postaction={decorate}
  ] 
  (100,0) -- (0,0) ;
  \filldraw[black] (0,0) circle[radius=1.2pt]
    node[below=2pt] {$x$};
  \filldraw[black] (100,0) circle[radius=1.2pt]
    node[below=2pt] {$y$};
\end{scope}
\end{tikzpicture}}} 
&= \sum_{k}\int\frac{iu_{k}\left(\mathbf{x}\right)u_{k}^{*}\left(\mathbf{y}\right)}{\omega-\varepsilon_{k}+i\eta_{k}0^{+}}\frac{e^{-i\omega\left(t_{x}-t_{y}\right)}\mathrm{d}\omega}{2\pi} 
\label{eq:internal_e} \\
\text{Coulomb line:} \quad 
\vcenter{\hbox{\tikzset{every picture/.style={line width=0.75pt}}
\begin{tikzpicture}[x=0.55pt,y=0.55pt,yscale=-1,baseline={([yshift=20pt]current bounding box.center)}]
\begin{scope}[shift={(150,100)}]
\draw[line width=1.2pt,
    decorate,
    decoration={snake, 
               amplitude=2pt,    
               segment length=8pt}] 
  (100,0) -- (0,0);
  \filldraw[black] (0,0) circle[radius=1.2pt]
    node[below=2pt] {$x$};
  \filldraw[black] (100,0) circle[radius=1.2pt]
    node[below=2pt] {$y$};
\end{scope}
\end{tikzpicture}}} 
&= \sum_{\mathbf{q}}\int\frac{i}{\left|\mathbf{q}\right|^{2}}\frac{e^{i\mathbf{q}\cdot\left(\mathbf{x}-\mathbf{y}\right)}}{V}\frac{e^{-i\omega\left(t_{x}-t_{y}\right)}\mathrm{d}\omega}{2\pi} 
\label{eq:coulomb_internal}\\
\text{massive mediator internal line:} \quad 
\vcenter{\hbox{\tikzset{every picture/.style={line width=0.75pt}}
\begin{tikzpicture}[x=0.55pt,y=0.55pt,yscale=-1,baseline={([yshift=20pt]current bounding box.center)}]
\begin{scope}[shift={(150,100)}]
\draw[line width=1.2pt,dashed] (100,0) -- (0,0);
\filldraw[black] (0,0) circle[radius=1.2pt]
    node[below=2pt] {$x$};
  \filldraw[black] (100,0) circle[radius=1.2pt]
    node[below=2pt] {$y$};
\end{scope}
\end{tikzpicture}}} 
&= \sum_{\mathbf{q}}\int\frac{-i}{\left|\mathbf{q}\right|^{2}+m_{A'}^{2}}\frac{e^{i\mathbf{q}\cdot\left(\mathbf{x}-\mathbf{y}\right)}}{V}\frac{e^{-i\omega\left(t_{x}-t_{y}\right)}\mathrm{d}\omega}{2\pi} 
\label{eq:internal_dark_photon}\\
\text{Coulomb vertex:} \quad 
\vcenter{\hbox{\tikzset{every picture/.style={line width=0.75pt}}
\begin{tikzpicture}[x=0.55pt,y=0.55pt,yscale=-1]
\begin{scope}[shift={(150,100)}]
  \filldraw[black] (0,0) circle[radius=1.3pt];
  \draw[line width=1.2pt,
    postaction={decorate,
    decoration={markings, mark=at position 0.57 with {\arrow{Triangle[scale=0.65, reversed]}}}}
  ] 
  (-41,-40) -- (0,0);
  \draw[line width=1.2pt,
    postaction={decorate,
    decoration={markings, mark=at position 0.57 with {\arrow{Triangle[scale=0.65, reversed]}}}}
  ] 
  (0,0) -- (-40,40);
  \draw[line width=1.2pt,
    decorate,
    decoration={snake, amplitude=2pt, segment length=8pt}
  ] 
  (0,0) -- (60,0);
  \node[below, inner sep=5pt] at (0,0) {$z$};
\end{scope}
\end{tikzpicture}}} 
&= -ie\int\mathrm{d}^{4}z 
\label{eq:coulomb}
\\
\text{electron-massive mediator vertex:} \quad 
\vcenter{\hbox{\tikzset{every picture/.style={line width=0.75pt}}
\begin{tikzpicture}[x=0.55pt,y=0.55pt,yscale=-1]
\begin{scope}[shift={(150,100)}]
  \filldraw[black] (0,0) circle[radius=1.3pt];
  \draw[line width=1.2pt,
    postaction={decorate,
    decoration={markings, mark=at position 0.5 with {\arrow{Triangle[scale=0.65, reversed]}}}}
  ] 
  (-41,-40) -- (0,0);
  \draw[line width=1.2pt,
    postaction={decorate,
    decoration={markings, mark=at position 0.57 with {\arrow{Triangle[scale=0.65, reversed]}}}}
  ] 
  (0,0) -- (-40,40);
  \draw[dashed,line width=1.2pt] (0,0) -- (60,0); 
  \node[below, inner sep=5pt] at (0,0) {$z$};
\end{scope}
\end{tikzpicture}}} 
&= ig_{e}\int\mathrm{d}^{4}z 
\label{eq:e-darkphoton vertex}
\\
\text{DM-massive mediator vertex:} \quad 
\vcenter{\hbox{\tikzset{every picture/.style={line width=0.75pt}}
\begin{tikzpicture}[x=0.55pt,y=0.55pt,yscale=-1]
\begin{scope}[shift={(150,100)}]
  \draw[gray!70,line width=1.2pt,
    postaction={decorate,
    decoration={markings, mark=at position 0.57 with {\arrow{Triangle[scale=0.65, reversed]}}}}
  ] 
  (-41,-40) -- (0,0);
  \draw[gray!70,line width=1.2pt,
    postaction={decorate,
    decoration={markings, mark=at position 0.57 with {\arrow{Triangle[scale=0.65, reversed]}}}}
  ] 
  (0,0) -- (-40,40);
  \draw[dashed,line width=1.2pt] (0,0) -- (60,0); 
  \node[below, inner sep=5pt] at (0,0) {$z$};
  \filldraw[black] (0,0) circle[radius=1.3pt];
\end{scope}
\end{tikzpicture}}} 
&= ig_{\chi}\int\mathrm{d}^{4}z 
\label{eq:DM-darkphoton vertex}
\end{align}
\end{subequations}In addition, a fermion loop brings an extra minus sign $\left(-1\right).$

\bibliographystyle{JHEP1}
\addcontentsline{toc}{section}{\refname}\bibliography{CuttingRules}

\end{document}